\documentclass{ws-procs11x85}

\usepackage{balance}

\def\Journal#1#2#3#4{{#1} {\bf #2}, #3 (#4)}

\makeindex
\begin{document}

\def\KP{K^+}
\def\KM{K^-}
\def\KZ{K^0}
\def\KS{{K^0_S}}
\def\KL{{K^0_L}}
\def\piZ{{\pi^0}}
\def\piP{{\pi^+}}
\def\piM{{\pi^-}}
\def\Kstar{{K^*}}
\def\KstarZ{{K^{*0}}}
\def\KstarP{{K^{*+}}}
\def\KPPM{\KP\piM}
\def\KSPZ{\KS\piZ}
\def\KSPP{\KS\piP}
\def\KPPZ{\KP\piZ}
\def\Jpsi{{J/\psi}}
\def\uubar{u\overline{u}{}}
\def\ddbar{d\overline{d}{}}
\def\ssbar{s\overline{s}{}}
\def\ccbar{c\overline{c}{}}
\def\qqbar{q\overline{q}{}}
\def\fbar{\overline{f}{}}
\def\Bbar{\overline{B}{}}
\def\Dbar{\overline{D}{}}
\def\Kbar{\overline{K}{}}
\def\pbar{\overline{p}{}}
\def\nubar{\overline{\nu}{}}
\def\sbar{\overline{s}{}}
\def\cbar{\overline{c}{}}
\def\bbar{\overline{b}{}}
\def\Xsbar{\overline{X}_s{}}
\def\epem{e^+e^-{}}
\def\mumu{\mu^+\mu^-{}}
\def\elel{\ell^+\ell^-{}}
\def\BBbar{B\Bbar}
\def\Bd{B_d^0{}}
\def\Bs{B_s^0{}}
\def\Bds{B_{d,s}^0{}}
\def\btosg{b\to sg}
\def\btosgamma{b\to s\gamma}
\def\bbartosbargamma{\bbar\to \sbar\gamma}
\def\btodgamma{b\to d\gamma}
\def\btosll{b\to s\elel}

\def\Btorhogamma{B\to \rho\gamma}
\def\Btoomegagamma{B\to \omega\gamma}
\def\BtoXsgamma{B\to X_s\gamma}
\def\BtoXdgamma{B\to X_d\gamma}
\def\BtoXsll{B\to X_s\elel}
\def\BtoXsee{B\to X_s \epem}
\def\BtoXsmumu{B\to X_s\mumu}
\def\BtoKll{B\to K\elel}
\def\BtoKee{B\to K\epem}
\def\BtoKmumu{B\to K\mumu}
\def\Xsll{X_s\elel}
\def\Xsee{X_s\epem}
\def\Xsmumu{X_s\mumu}
\def\Kll{K\elel}
\def\Kstarll{K^*\elel}
\def\BtoKstarll{B\to K^*\elel}
\def\BtoKstaree{B\to K^*\epem}
\def\BtoKstarmumu{B\to K^*\mumu}
\def\BtoKorKstarll{B\to K^{(*)}\elel}
\def\BtoKstargamma{B\to K^*\gamma}
\def\BbartoKstarbargamma{\Bbar\to \Kbar^*\gamma}
\def\BtoKZG{B^0\to K^{*0}\gamma}
\def\BtoKPG{B^+\to K^{*+}\gamma}
\def\BtoKstarZG{B^0\to K^{*0}\gamma}
\def\BtoKstarPG{B^+\to K^{*+}\gamma}
\def\BtoKXgamma{B\to K_X\gamma}
\def\BtoVgamma{B\to V\gamma}
\def\BtoKstarzerogamma{B^0\to K^{*0}\gamma}
\def\BtoKstarplusgamma{B^+\to K^{*+}\gamma}
\def\Btorhozerogamma{B^0\to \rho^0\gamma}
\def\Btorhoplusgamma{B^+\to \rho^+\gamma}

\def\BtoKpi{B\to K\pi}
\def\BtophiK{B\to\phi K}
\def\Btopilnu{B\to\pi\ell\nu}
\def\Btorholnu{B\to\rho\ell\nu}
\def\Btoomegalnu{B\to\omega\ell\nu}
\def\Btopipi{B\to\pi\pi}
\def\BtoDspi{B\to D_s\pi}
\def\ps{\mbox{~ps}}
\def\um{\mbox{~}\mu\mbox{m}}
\def\mm{\mbox{~mm}}
\def\cm{\mbox{~cm}}

\def\lumiunit{\mbox{~cm}^{-2}\mbox{s}^{-1}}

\def\nb{\mbox{~nb}}
\def\nbinv{\mbox{~nb}^{-1}}
\def\pbinv{\mbox{~pb}^{-1}}
\def\fbinv{\mbox{~fb}^{-1}}
\def\abinv{\mbox{~ab}^{-1}}

\def\GeV{\mbox{~GeV}}
\def\GeVc{\mbox{~GeV}/c}
\def\GeVcc{\mbox{~GeV}/c^2}
\def\MeV{\mbox{~MeV}}
\def\MeVc{\mbox{~MeV}/c}
\def\MeVcc{\mbox{~MeV}/c^2}
\def\Vud{V_{ud}}
\def\Vus{V_{us}}
\def\Vub{V_{ub}}
\def\Vcd{V_{cd}}
\def\Vcs{V_{cs}}
\def\Vcb{V_{cb}}
\def\Vtd{V_{td}}
\def\Vts{V_{ts}}
\def\Vtb{V_{tb}}

\def\Order#1{{\cal O}(#1)}
\def\Br{{\cal B}}
\def\Lumi{{\cal L}}

\def\Mbc{M_{\rm bc}}
\def\DeltaE{\Delta{E}}
\def\Ebeam{E^*_{\rm beam}{}}
\def\EB{E_B^*{}}
\def\pB{p_B^*{}}
\def\Emiss{E_{\rm miss}}
\def\Pmiss{P_{\rm miss}}
\def\pCM{p^{\rm CM}}

\def\Mll{M(\elel)}
\def\MKpi{M(K\pi)}

\def\shat{\hat{s}}
\def\Cseven{C_7^{\rm eff}}
\def\Cnine{C_9^{\rm eff}}
\def\Cten{C_{10}^{\rm eff}}

\def\Egamma{E_\gamma}
\def\Egammamin{E_\gamma^{\rm min}}
\def\Acp{A_{CP}}
\def\Deltapz{\Delta_{+0}}
\def\tauBratio{\tau_{B^+}/\tau_{B^0}}

\def\MXs{M(X_s)}
\def\AFB{A_{FB}}

\def\costhetahel{{\cos\theta_{\rm hel}}}

\def\redto{\red\Rightarrow}

\def\PM#1#2{\,^{+#1}_{-#2}{}}
\def\EM#1{\times10^{-#1}}

\def\etal{\textit{et al.}}
\def\Journal#1#2#3#4{{#1} {\bf #2}, #3 (#4)} 
\def\NCA{\em Nuovo Cimento}                      
\def\NIMA{{\em Nucl. Instrum. Meth.} A}
\def\NPB{{\em Nucl. Phys.} B}
\def\PLB{{\em Phys. Lett.} B}
\def\PRL{\em Phys. Rev. Lett.}
\def\PRD{{\em Phys. Rev.} D}
\def\ZPC{{\em Z. Phys.} C}
\def\EPJD{{\em Eur. Phys. J. direct} C}
\def\EPJC{{\em Eur. Phys. J.} C}

\def\CITE#1#2{\rlap#1\,\cite{#2}}

\title{RADIATIVE AND ELECTROWEAK RARE {\boldmath $B$} DECAYS}

\author{M. NAKAO}

\address{KEK, High Energy Accelerator Research Organization,
1--1 Oho, Tsukuba, Ibaraki 305--0801, JAPAN\\E-mail: mikihiko.nakao@kek.jp}



\twocolumn[\maketitle\abstract{ This report summarizes the latest
experimental results on radiative and electroweak rare $B$ meson decays.
These rare decay processes proceed through the Flavor-Changing-Neutral-Current 
processes, and thus are sensitive to the postulated new particles in
the theories beyond the Standard Model.  Experiments at $e^+e^-$
colliders, Belle, BaBar and CLEO, have been playing the dominant role,
while the CDF and D0 experiments have just started to provide new
results from Tevatron Run-II.  The most significant achievement is the
first observation of the decay $\BtoKstarll$, which opens a new window
to search for new physics in $B$ meson decays. }]


\baselineskip=13.07pt
\section{Introduction}

Rare $B$ meson decays that include a photon or a lepton pair in the
final state have been the most reliable window---besides the
Cabibbo-Kobayashi-Maskawa (CKM) unitarity triangle---to understand the
framework of the Standard Model (SM) using the rich sample of $B$
decays, and to search for physics beyond the SM.  Belle has just
reported that the CP-violating phase in $B\to\phi\KS$ may 
deviate largely from the SM expectation measured using the $B\to J/\psi\KS$ and
related modes\CITE.{bib:belle-phiks}  The former is the $b\to s\sbar s$
transition which proceeds presumably through the loop (penguin) diagram for
the $b\to s$ Flavor-Changing-Neutral-Current (FCNC) process, while the
latter is the $b\to c\cbar s$ transition which is dominated by the tree diagram
and is unlikely to be interfered with by new physics with a large effect.
It is therefore an urgent question whether we can also find a similar
deviation from the SM in any other related $b\to s$ transitions using
the large samples of $\epem\to\Upsilon(4S)\to\BBbar$ data available from
two $B$-factories, Belle and BaBar, in order to investigate the nature
of the possible new physics signal.

\begin{figure}[b]
  \center
  \psfig{figure=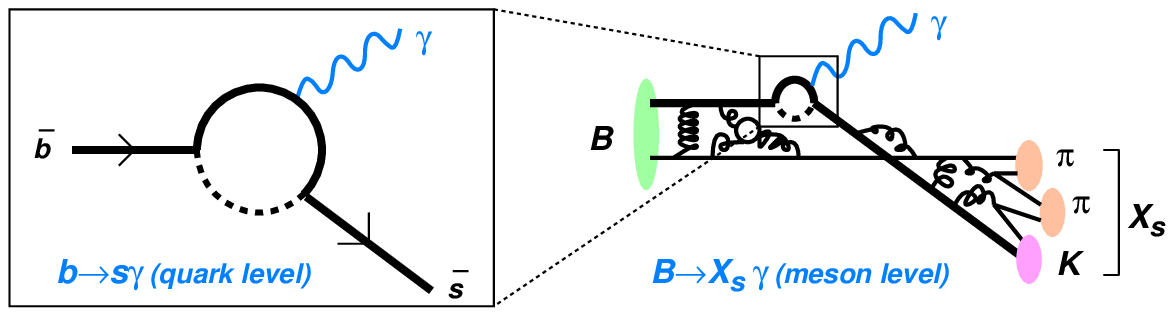,width=8truecm}
  \caption{$\btosgamma$ and $\BtoXsgamma$.}
  \label{fig:btosgamma}
\end{figure}

Radiative $B$ decays with a high energy photon in the final state are a
unique probe to explore inside the $B$ meson.  In the SM, the high
energy photon is radiated through FCNC processes, $\btosgamma$ and
$\btodgamma$.  These transitions are forbidden at the tree level and
only proceed via penguin loops formed by a virtual top quark and a $W$
boson, or other higher order diagrams.  The loop diagram can also be
formed by postulated heavy particles if they exist, and is therefore
sensitive to physics beyond the SM.  The $\btosgamma$ decay rate is
large enough to have been measured already by CLEO\cite{bib:xsgam-cleo}
and ALEPH\CITE,{bib:xsgam-aleph} and then by Belle\cite{bib:xsgam-belle}
and BaBar\CITE.{bib:xsgam-babar-full,bib:xsgam-babar-semi}  As
illustrated in Fig.~\ref{fig:btosgamma}, the $\btosgamma$ transition at
the quark level can be studied by performing an inclusive measurement
for $\BtoXsgamma$, where $X_s$ is an inclusive state with a strangeness
$S=\pm1$.  The photon energy spectrum, which can be characterized by its
mean energy and moments, provides a useful constraint to the heavy quark
effective theory that essentially helps to reduce the uncertainties in
the inclusive semi-leptonic $B$ decay rates and hence the extraction of
$|\Vcb|$ and $|\Vub|$.
%
%
In contrast to the inclusive studies, exclusive decay modes such as
$\BtoKstargamma$ are experimentally much easier to measure and have been
extensively explored.  However, one has to always consider large model
dependent hadronic uncertainties to compare the results with the SM.
Such uncertainties largely cancel by searching for CP- and isospin
asymmetries.  Though the $\btodgamma$ transition is suppressed by a CKM
factor $|\Vtd/\Vts|^2\sim\Order{10^{-2}}$ with respect to $\btosgamma$,
searches are still being pursued for this exclusive decay channel.

Electroweak rare $B$ decays proceeding through a similar FCNC process $\btosll$
($\ell=e,\mu$) involves a virtual photon or weak boson, and has
sensitivities to new physics that are not covered by $\btosgamma$.  This
process is suppressed with respect to $\btosgamma$ by an additional
$\alpha_{\rm em}$ factor that has made it inaccessible before
the $B$-factories.  Having two leptons in the final state, one can measure
the dependence on the momentum transfer squared $q^2(=\Mll^2)$ of the
virtual $\gamma/Z$.  Furthermore, measurement of the forward-backward
asymmetry of the lepton decay angle will be a unique probe in this
electroweak process, with a small theoretical uncertainty even in the
exclusive decay $\BtoKstarll$.  The pure weak process, $b\to s\nu\nubar$ is
experimentally extremely difficult.

Pure leptonic decay $B^0_{d,s}\to\elel$ is based on the same quark
diagram as $b\to(d,s)\elel$, and hence has a similar sensitivity to new
physics.  The SM expected branching fractions are beyond the current
experimental reach, but new physics may dramatically enhance the decay
rate, especially for $B_s\to\mumu$, for which the Tevatron experiments
have just restarted to provide new information.  The charged counter
part, $B^+\to\tau^+\nu$ and $B^+\to\ell^+\nu$, are tree level processes.
These decays have not been observed yet because of the very small
branching fractions due to the GIM suppression mechanism and the
experimental difficulty due to the missing neutrino.

In this report, the latest results on radiative (Sec.~2), electroweak
(Sec.~3) and pure leptonic (Sec.~4) $B$ decays are reviewed.
Belle has analyzed up to $140\fbinv$ corresponding to 152 million
$\BBbar$ pairs, while BaBar has analyzed up to $113\fbinv$ corresponding
to 123 million $\BBbar$ pairs.  The first results from the Tevatron Run-II
data from CDF and D0 are also included.  Finally Sec.~5 concludes this
report.


\section{Radiative $B$ Decays}


\subsection{Inclusive $\BtoXsgamma$ Branching Fraction}

\begin{figure}
  \center
  \psfig{figure=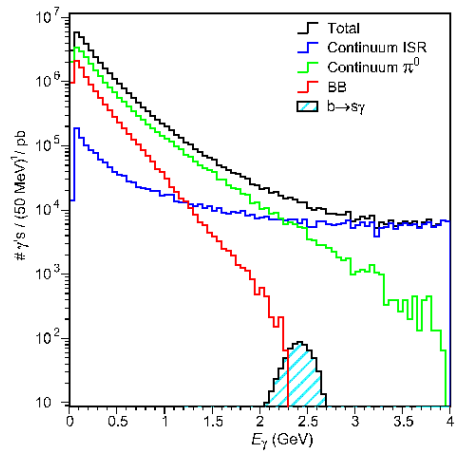,width=6truecm}
  \caption{Expected photon energy distribution for $\BtoXsgamma$ signal
  and various background sources.}
\vspace{-0.4cm}
  \label{fig:egam}
\end{figure}

Due to the two-body decay nature of the quark level process of
$\btosgamma$, the photon energy spectrum of $\BtoXsgamma$ has a peak
around half of the $b$ quark mass.  This peak is the signature of
the fully inclusive $\BtoXsgamma$ measurement.  On top of this signal,
there are huge background sources as shown in Fig.~\ref{fig:egam}.  The
largest contribution is from the continuum process $\epem\to\qqbar$
($q=u,d,s,c$) in which copious $\piZ\to\gamma\gamma$ and
$\eta\to\gamma\gamma$ are the sources of high energy photons, and the
initial-state radiation process $\epem\to\qqbar\gamma$.  These continuum
backgrounds are reliably subtracted by using the off-resonance data
sample taken slightly below the $\Upsilon(4S)$ resonance.  The
background from $B$ decay is also significant, especially for lower
photon energies.  To estimate and subtract the $B$ decay background, one
has to largely rely on the Monte Carlo simulation.  As the signal rate
rapidly decreases and the background rate rapidly increases towards 
lower photon energies, it is inevitable that one requires a minimum photon energy
($\Egammamin$) and extrapolates the spectrum below $\Egammamin$ to
obtain the total branching fraction.

An alternative semi-inclusive method is to sum up all the possible fully
reconstructed $X_s\gamma$ final states, where $X_s$ is formed from one
kaon and up to four pions.  In this case, one can require the kinematic
constraints on the beam-energy constrained mass
$\Mbc=\sqrt{\Ebeam^2-\pB^2}$ (also denoted as the beam-energy
substituted mass $M_{ES}$) and $\DeltaE=\EB-\Ebeam$, using the beam
energy $\Ebeam$ and fully reconstructed momentum $\pB$ and energy $\EB$
of the $B$ candidate in the center-of-mass (CM) frame. Therefore
the large backgrounds can be reduced at the cost of introducing an additional
error due to the model dependent hadronization uncertainties.

\begin{figure}
  \center
  \psfig{figure=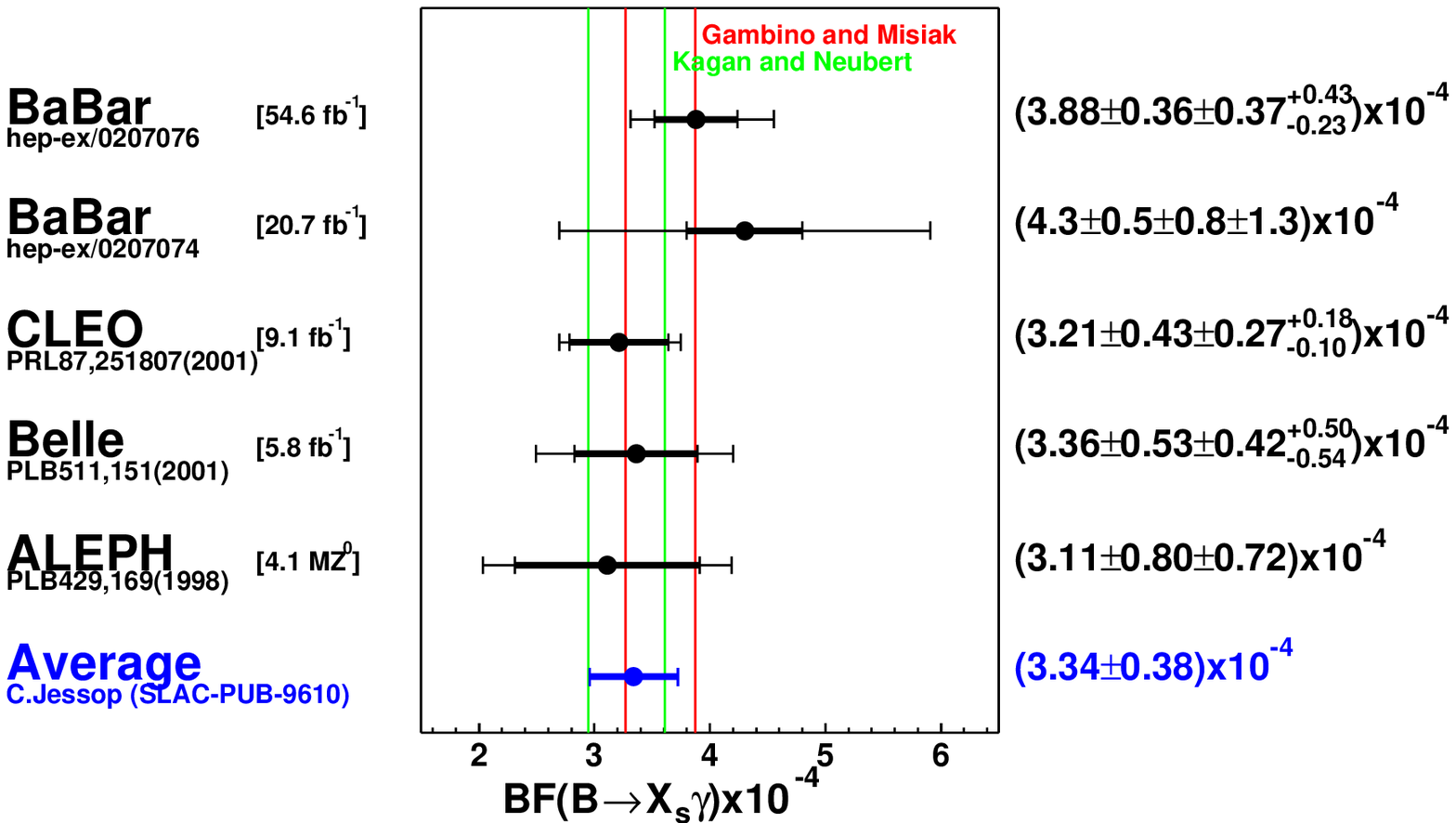,width=8truecm}
  \caption{$\BtoXsgamma$ branching fraction measurements.}
\vspace{-0.4cm}
  \label{fig:xsgam}
\end{figure}

So far, CLEO\cite{bib:xsgam-cleo} and BaBar\cite{bib:xsgam-babar-full}
have performed the fully inclusive measurement, and
Belle\cite{bib:xsgam-belle} and BaBar\cite{bib:xsgam-babar-semi} have
performed the semi-inclusive measurement.  Figure~\ref{fig:xsgam}
summarizes these results, together with the measurement performed by
ALEPH\CITE.{bib:xsgam-aleph}  CLEO has applied the lowest $\Egammamin$
of $2.0\GeV$ and has the smallest error, while BaBar requires
$\Egammamin=2.1\GeV$, and Belle requires $\MXs<2.1\GeV$ which is roughly 
equivalent to $\Egammamin \sim 2.25\GeV$.

The average of the five measurements, including the two from BaBar with an
overlapping data set, has been calculated taking into account the
correlation of the systematic and theory errors, as\cite{bib:xsgam-avg}
\begin{equation}
\Br(\BtoXsgamma)=(3.34\pm0.38)\times10^{-4}.
\end{equation}
The latest SM calculation\cite{bib:xsgam-sm-latest} predicts
$\Br(\BtoXsgamma)=(3.57\pm0.30)\times10^{-4}$, in very good agreement
with the world average.  The prediction fully includes up to
Next-to-Leading-Order QCD corrections\CITE.{bib:xsgam-sm-nlo}

This result can be used to constrain new physics
hypotheses\CITE.{bib:xsgam-sm-latest,bib:xsgam-2hdm,bib:xsgam-bsm}  For
example, any new physics that has only a constructive interference with
the SM amplitude is strongly constrained.  The Type-II charged Higgs boson
is one such example, and its mass has to be greater than $350\GeV$
if there is no other destructive
amplitude\CITE.{bib:xsgam-sm-latest,bib:xsgam-2hdm}  Many SUSY models
can, however, have also a destructive amplitude that may cancel the
constructive part.  The decay amplitude is usually written down using
the effective Hamiltonian with Wilson coefficients for the relevant
operators.  The $\BtoXsgamma$ result constrains the magnitude of the
$C_7$ Wilson coefficient, that can be a useful measure of a possible
deviation from the SM, and also is an input parameter to the constraints
provided by other measurements.

In order to further improve the measurement, it is necessary to lower
the minimum photon energy.  The latest Belle and BaBar data samples,
with the largest off-resonance data size, have yet to be analyzed.  A
new effort to significantly reduce the theory error by including the
Next-to-Next-to-Leading-Order QCD correction has also been started.


\subsection{Exclusive $\BtoKstargamma$}

\begin{figure}
  \center
  \psfig{figure=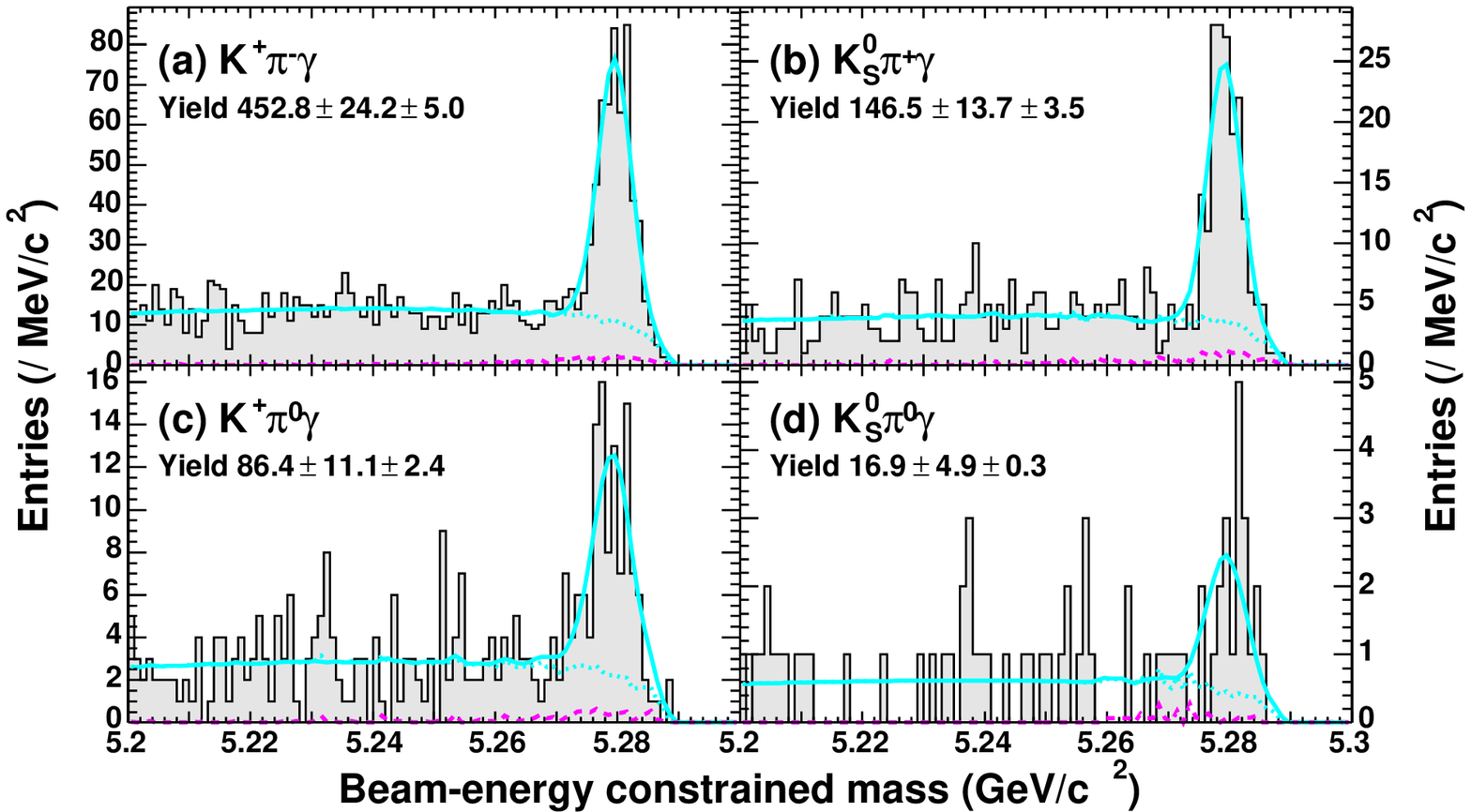,width=8truecm}
  \caption{$\BtoKstargamma$ signal from Belle.}
  \label{fig:belle-kstargam}
\end{figure}

The measurement of the $\BtoKstargamma$ exclusive branching fraction is
straightforward, since one can use the $\Mbc$, $\DeltaE$ and $K^*$ mass
constraints.  ($K^*$ denotes $K^*(892)$ throughout this report.)  The
latest Belle measurement (Fig.~\ref{fig:belle-kstargam}) uses
$78\fbinv$ data, with a total error of much less than 10\% for each of the
$B^0$ and $B^+$ decays.  The results from CLEO\CITE,{bib:kstargam-cleo}
BaBar\cite{bib:kstargam-babar} and Belle\cite{bib:kstargam-belle} are in
good agreement and are listed in Table~\ref{tbl:kstargam}. The world averages
are calculated as
\begin{equation}
  \Br(\BtoKstarZG)=(4.17\pm0.23)\times10^{-5},
\end{equation}
\begin{equation}
  \Br(\BtoKstarPG)=(4.18\pm0.32)\times10^{-5}.
\end{equation}
The corresponding theoretically predicted branching fraction is about
$(7\pm2)\times10^{-5}$, higher than the measurement with a large
uncertainty\CITE.{bib:kstargam-sm}  As the $\btosgamma$ transition is well
understood by the inclusive measurement, we consider the deviation is
due to the ambiguous hadronic form factor, for which the light-cone QCD
sum rule result of $F_7^{B\to K^*}(0)=0.38\pm0.05$ is used.  However,
a recent lattice-QCD calculation\cite{bib:kstargam-becirevic} is
suggesting that the expected form-factor is as small as $F_7^{B\to
K^*}(0)=0.25\pm0.04$ and is consistent with the value of $F_7^{B\to
K^*}(0)=0.27\pm0.04$ extracted from the measured branching fraction.

\begin{table}
\begin{center}
\caption{$B\to K^*\gamma$ branching fractions\label{tbl:kstargam}}
\begin{tabular}{ccc}
\hline
 & $B^0\to K^{*0}\gamma$
 & $B^+\to K^{*+}\gamma$ \\
 & $[\times10^{-5}]$ &  $[\times10^{-5}]$ \\
\hline
CLEO  & $4.55\pm0.70\pm0.34$ & $3.76\pm0.86\pm0.28$ \\
BaBar & $4.23\pm0.40\pm0.22$ & $3.83\pm0.62\pm0.22$ \\
Belle & $4.09\pm0.21\pm0.19$ & $4.40\pm0.33\pm0.24$ \\
\hline
\end{tabular}
\end{center}
\end{table}

A better approach to exploit the $\BtoKstargamma$ branching fraction
measurements is to consider isospin
asymmetry\CITE.{bib:kstargam-isospin}  A small difference in the
branching fractions between $\BtoKstarZG$ and $\BtoKstarPG$ tells us the
sign of the combination of the Wilson coefficients, $C_6/C_7$.  Belle
has taken into account the correlated systematic errors and performed a
measurement as
\begin{equation}
\begin{array}{lll}
\Deltapz&\equiv&{\displaystyle
                {(\tauBratio)\Br(\BtoKstarZG)-\Br(\BtoKstarPG) \over
                 (\tauBratio)\Br(\BtoKstarZG)+\Br(\BtoKstarPG)}}\\[12pt]
&=&(+0.003\pm0.045\pm0.018),
\end{array}
\end{equation}
which is consistent with zero and one cannot tell whether the SM prediction
($\Deltapz>0$) is correct yet.  Here, the lifetime ratio
$\tauBratio=1.083\pm0.017$ is used, and the $B^0$ to $B^+$ production
ratio is assumed to be unity.  The latter is measured to be
$f_0/f_+=1.072\pm0.057$ and is a source of an additional systematic
error.

\subsection{Other Exclusive Radiative Decays}

The dominant radiative decay channel $\BtoKstargamma$ covers only 12.5\%
of the total $\BtoXsgamma$ branching fraction, and the rest has to be
accounted for by decays with higher resonances or multi-body decays.
Knowledge of these decay modes will eventually be useful to reduce the
systematic error of the inclusive measurement.  Some of the decays have a
particular property that is useful to search for new physics.  As an
example, the decay channel $B^0\to K_1(1270)^0\gamma\to\KS\rho^0\gamma$
will be useful to measure the time-dependent CP-asymmetry;\cite{bib:m0gam-atwood} 
while another such measurement is
experimentally challenging: $B^0\to K^{*0}\gamma\to\KS\piZ\gamma$
using the detached $\KS\to\piP\piM$ decay vertex.  Another example is to
use the decay $B^+\to K_1(1400)^+\gamma\to \KS\piP\piZ$ for a photon
polarization measurement\CITE.{bib:kpipigam-gronau}

\begin{figure}
  \center
  \psfig{figure=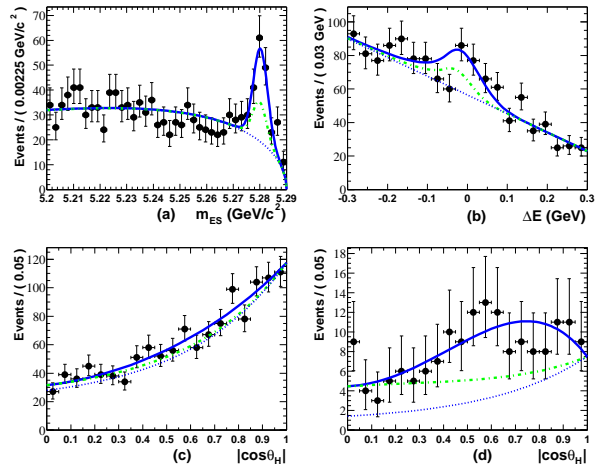,width=8truecm}
  \caption{$B^0\to K_2^*(1400)^0\gamma$ signal by BaBar.}
  \label{fig:babar-k2stgam}
\end{figure}

The $B\to K_2^*(1430)\gamma$ decay mode is unique since the $K_2^*(1430)$
decays into a $K\pi$ combination, while many other resonances have very
small or no decay width to $K\pi$.  After measurements by
CLEO\cite{bib:kstargam-cleo} and Belle\CITE,{bib:kxgam-belle} BaBar has
also performed a new measurement\cite{bib:k2gam-babar}
(Fig.~\ref{fig:babar-k2stgam}).  Branching fractions are listed in
Table~\ref{tbl:k2stargam}.  The results are in agreement with the SM
predictions\CITE,{bib:veseli-olsson} for example,
$(17.3\pm8.0)\times10^{-6}$.

\begin{table}
\begin{center}
\caption{$B\to K_2^*(1430)\gamma$ branching fractions.}
\label{tbl:k2stargam}
\begin{tabular}{ccc}
\hline
 & $B^0\to K_2^*(1430)^0\gamma$
 & $B^+\to K_2^*(1430)^+\gamma$ \\
 & $[\times10^{-6}]$ &  $[\times10^{-6}]$ \\
\hline
CLEO & \multicolumn{2}{c}{$16.6\PM{5.9}{5.3}\pm1.3$} \\
Belle & $13\pm5\pm1$ & --- \\
BaBar & $12.2\pm2.5\pm1.1$ & $14.4\pm4.0\pm1.3$ \\
\hline
\end{tabular}
\end{center}
\end{table}

\begin{figure}
  \center
  \psfig{figure=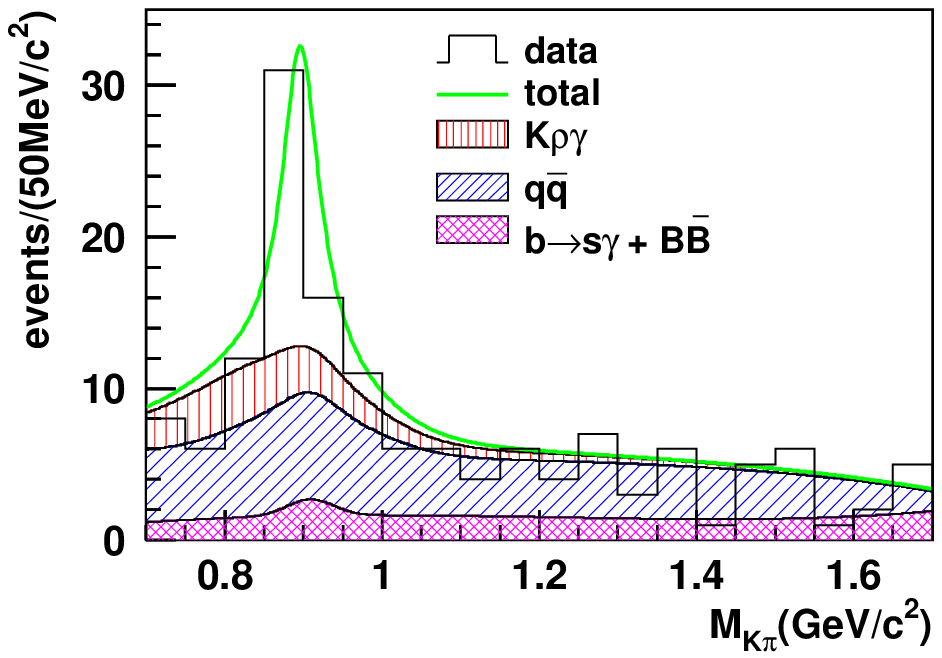,width=3.9truecm}
  \psfig{figure=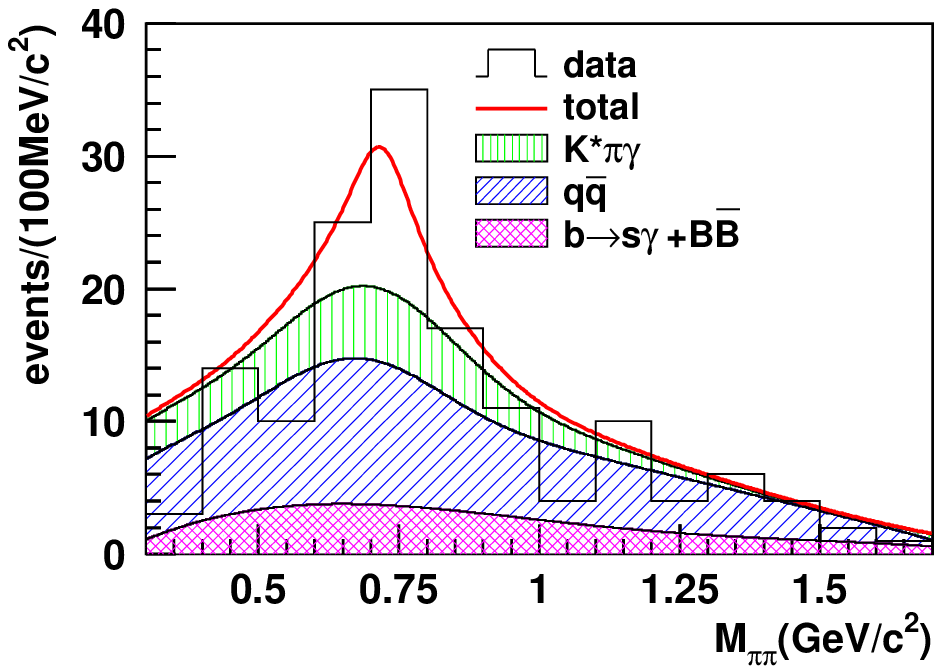,width=3.9truecm}
  \caption{$B^+\to K^{*0}\pi^+\gamma$ and $B\to K^+\rho^0\gamma$ from Belle.}
  \label{fig:belle-kxgam}
\end{figure}

\begin{figure}
  \center
  \psfig{figure=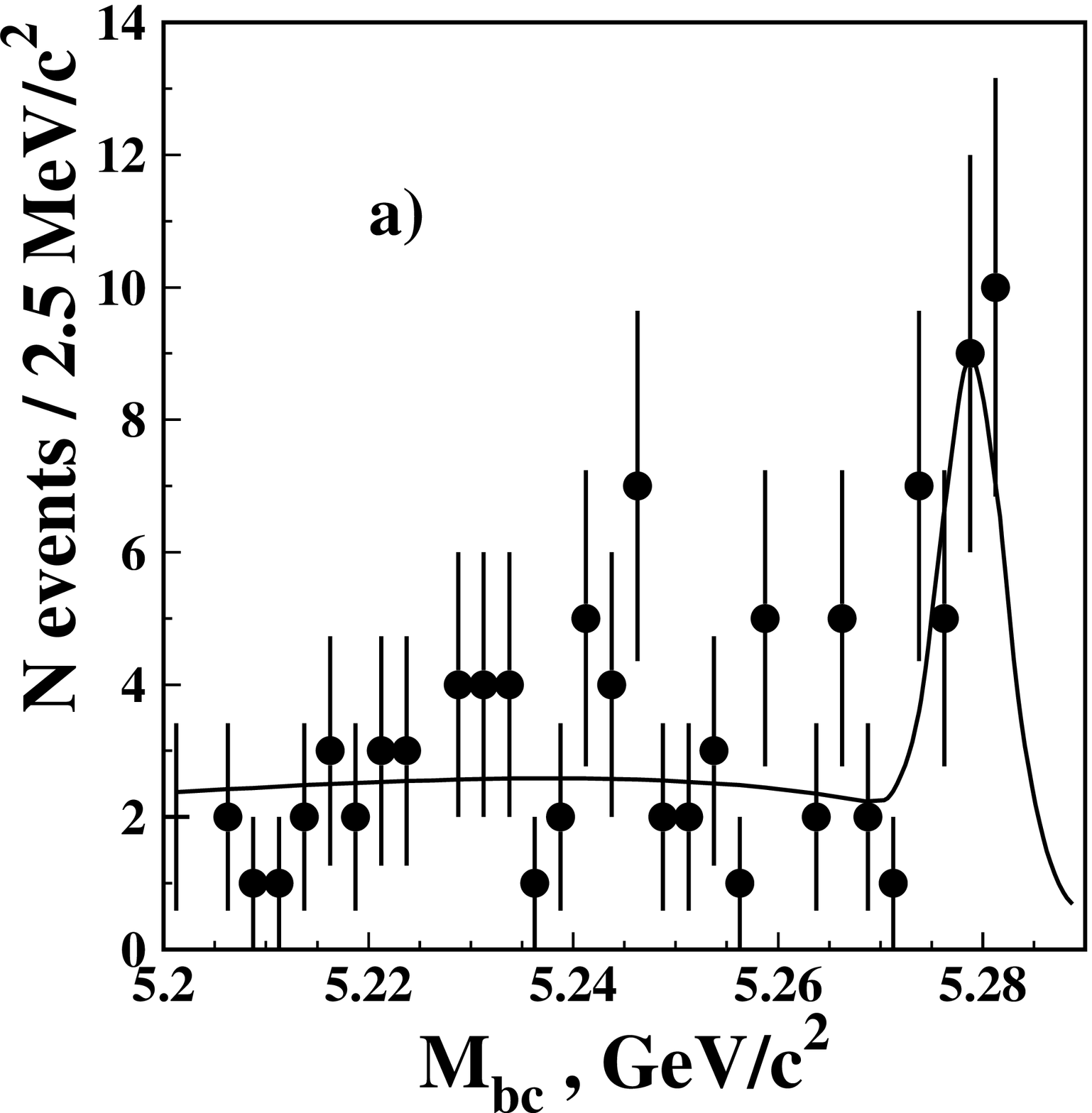,width=3.9truecm}
  \psfig{figure=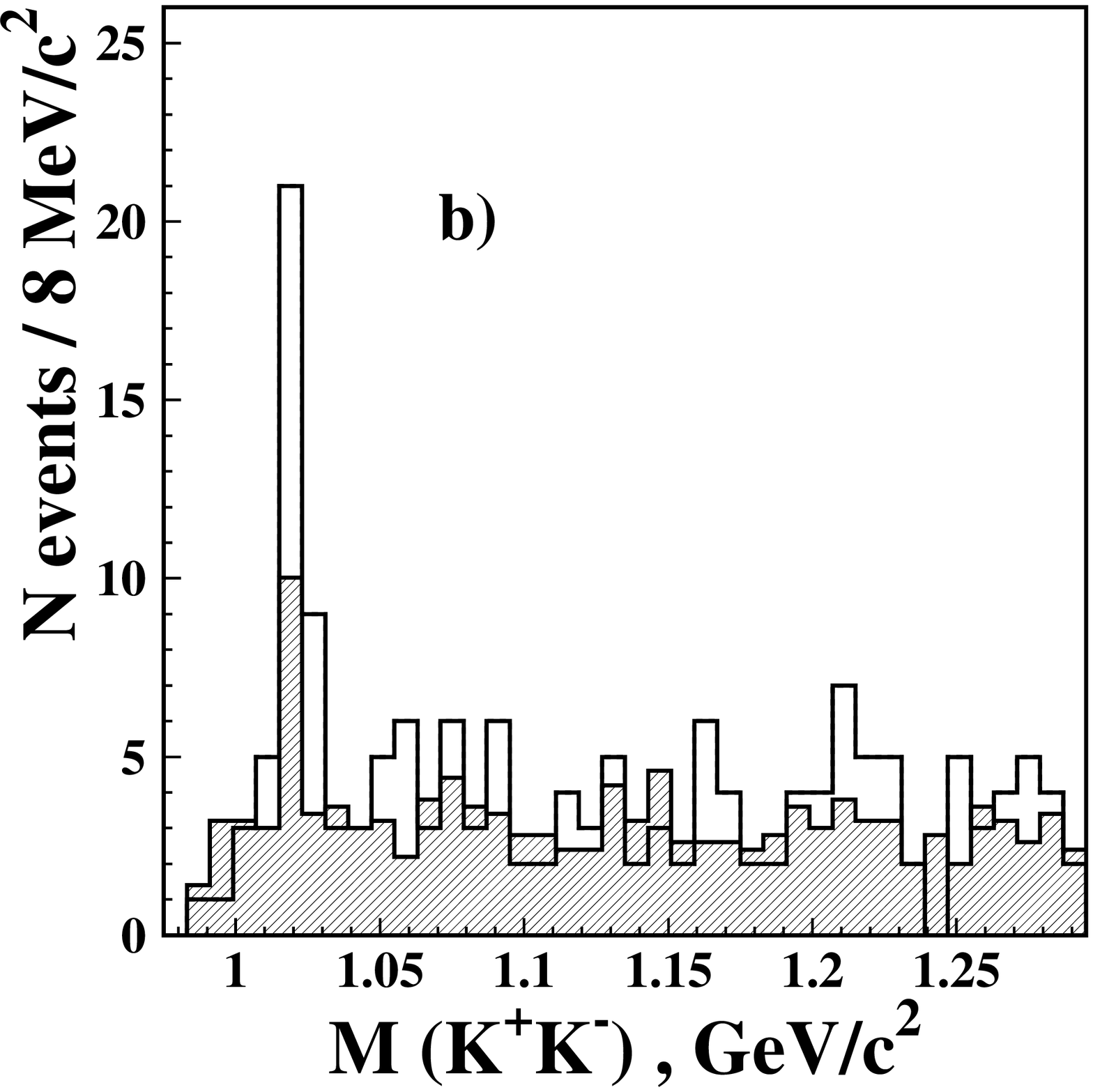,width=3.9truecm}
  \caption{$B^+\to K\phi\gamma$ from Belle.}
  \label{fig:belle-kphigam}
\end{figure}

Belle has extended the analysis into multi-body decay
channels\CITE.{bib:kxgam-belle}  Using $29\fbinv$ data, the decay
$B^+\to \KP\piP\piM\gamma$ is measured to have a branching fraction of
$(24\pm5\PM{4}{2})\times10^{-6}$ for $M(K\pi\pi)<2.4\GeV$.  The decay is
dominated by $K^{*0}\piP\gamma$ and $\KP\rho^0\gamma$ final states that
overlap each other as shown in Fig.~\ref{fig:belle-kxgam}.  At
this moment, it is not possible to disentangle resonant states that
decay into $K^*\pi$ or $K\rho$, such as $K_1(1270)$, $K_1(1400)$,
$K^*(1650)$, and so on.  A clear $B^+\to K^+\phi\gamma$ ($5.5\sigma$)
signal was recently observed by Belle with $90\fbinv$ data
(Fig.~\ref{fig:belle-kphigam}), together with a $3.3\sigma$ evidence for
$B^0\to\KS\phi\gamma$.  There is no known $K\phi$ resonant state.  This
is the first example of a $s\sbar s\gamma$ final state.  Branching
fractions are measured to be\cite{bib:kphigam-belle}
\begin{equation}
\begin{array}{rcl}\displaystyle
\Br(B^+\to K^+\phi\gamma)&=&(3.4\pm0.9\pm0.4)\times10^{-6}\\
\Br(B^0\to K^0\phi\gamma)&=&(4.6\pm2.4\pm0.4)\times10^{-6}\\
                         &<&8.3\times10^{-6} \mbox{~~~(90\%~CL)}
\end{array}
\end{equation}
With more data, one can perform a time-dependent CP-asymmetry
measurement with the $\KS\phi\gamma$ decay channel.

Radiative decays with baryons in the final state have been searched for
by CLEO\CITE,{bib:Lpgam-cleo} in the $B^-\to\Lambda\pbar\gamma$ channel
for photon energies greater than $2\GeV$.  The analysis is also
sensitive to $B^-\to\Sigma^0\pbar\gamma$ with a slightly shifted
$\DeltaE$ signal window due to the missing soft photon in
$\Sigma^0\to\Lambda\gamma$.  Upper limits are given as
\begin{equation}
\begin{array}{rcl}\displaystyle
\Br(B^-\to \Lambda\pbar\gamma)+0.3\Br(B^-\to\Sigma^0\pbar\gamma) &<& 3.3\times10^{-6}\\
\Br(B^-\to \Sigma^0\pbar\gamma)+0.4\Br(B^-\to\Lambda\pbar\gamma) &<& 6.4\times10^{-6}.\\
\end{array}
\end{equation}
Considering isospin and other resonances such as $N(1232)$, an upper
limit on baryonic radiative decay is obtained to be less than
$3.8\times10^{-5}$, or 13\% of the total $\BtoXsgamma$ branching
fraction.

In summary, $(35\pm6)\%$ of the total $\BtoXsgamma$ is measured to be
one of $\BtoKstargamma$ (12.5\%), $B\to K_2^*(1430)\gamma$ (4\% after
excluding $K\pi\pi\gamma$), $B\to K^*\pi\gamma$ (9\%), $B\to
K\rho\gamma$ (9\%) or $B\to K\phi\gamma$ (1\%). The remaining $(65\pm6)\%$
would be accounted for by decays with multi-body final states, baryonic
decays, modes with $\eta$ and $\eta'$, multi-kaon final states other
than $K\phi\gamma$ or in the large $X_s$ mass range.


\subsection{Search for Direct CP-asymmetry}

Direct CP-asymmetry in $\BtoXsgamma$ is predicted to be 0.6\% in the SM
with a small error\CITE.{bib:acpxsgam-sm-orig,bib:acpxsgam-kagan}  This
is contrary to the other hadronic decay channels with the $b\to s$
transition, for which usually larger SM CP-asymmetries are predicted,
however, with large uncertainties.  Although such a small SM asymmetry
is beyond the sensitivity of the current $B$-factories, many extensions
to the SM predict that it is possible to produce a large CP-asymmetry
greater than 10\%\CITE.{bib:acpxsgam-kagan,bib:acpxsgam-bsm}  A large CP-asymmetry 
will be a clear sign of new physics.

There has been only one measurement by CLEO\cite{bib:acpxsgam-cleo} to
search for the direct CP-asymmetry of the radiative decays, which is
sensitive also to $\BtoXdgamma$.  The result is expressed as
\begin{equation}
\begin{array}{l}\displaystyle
0.965\Acp(\BtoXsgamma)+0.02\Acp(\BtoXdgamma)\\
               =(-0.079\pm0.108\pm0.022)\times(1\pm0.030).
\end{array}
\end{equation}
The SM predicts that $\BtoXdgamma$ has a much larger $\Acp$ with an
opposite sign to that of $\BtoXsgamma$.

\begin{figure}
  \center
  \psfig{figure=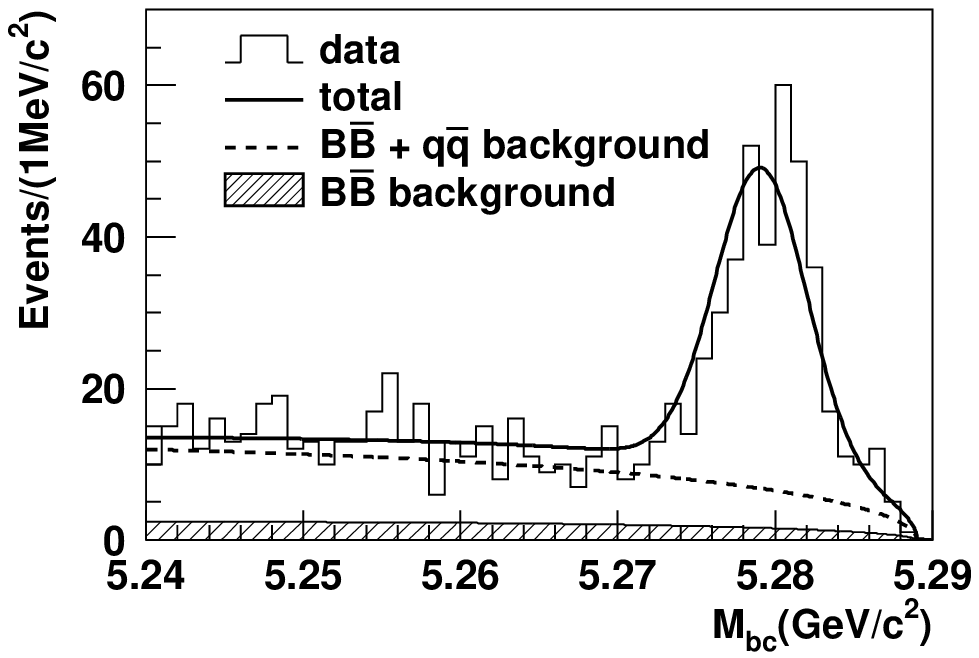,width=3.9truecm}
  \psfig{figure=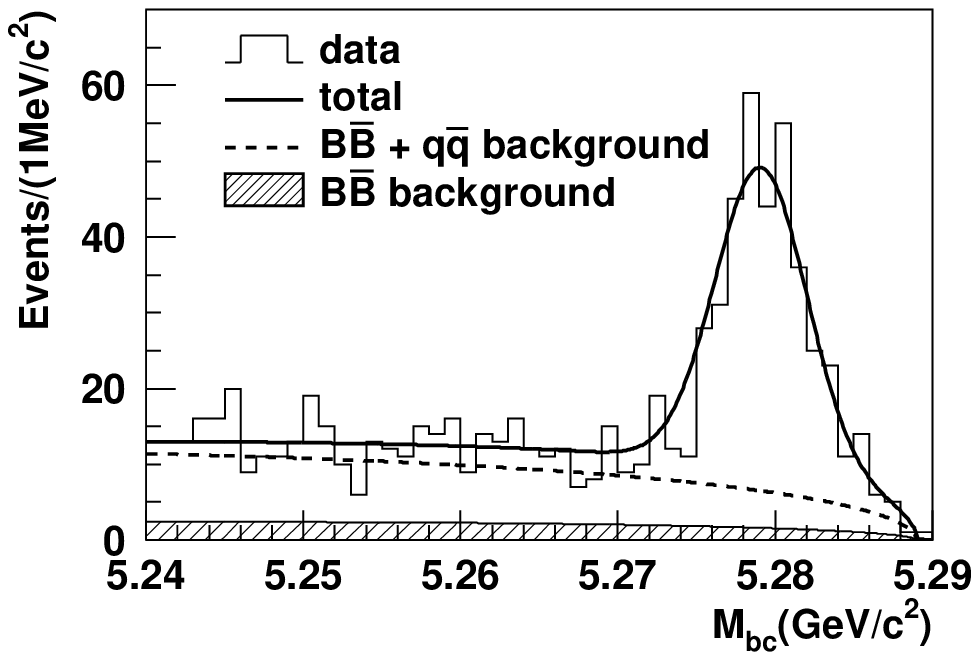,width=3.9truecm}
  \psfig{figure=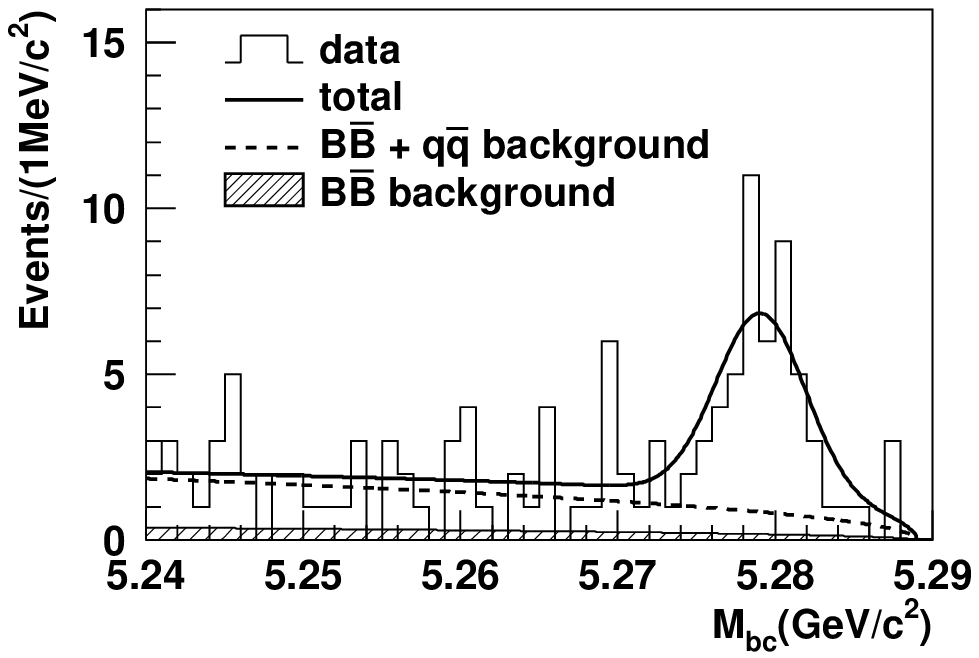,width=3.9truecm}
  \caption{$\Bbar$-tagged (top-left),
           $B$-tagged (top-right) and ambiguous (bottom) $\BtoXsgamma$
           signal from Belle.}
  \label{fig:belle-acpxsgam}
\end{figure}

A new $\Acp(\BtoXsgamma)$ measurement performed by
Belle\cite{bib:acpxsgam-belle} uses a similar technique to CLEO's, summing
up the exclusive modes of one kaon plus up to four pions.  In addition,
modes with three kaon plus up to one pion are included.  Belle's result
eliminates $\BtoXdgamma$ by exploiting particle identification devices
for the tagged hadronic recoil system.  CLEO requires $\Egammamin=2.2\GeV$
while Belle require $\MXs<2.1\GeV$ which roughly corresponds to
$\Egammamin \sim 2.25\GeV$. Events are self-tagged as $B$ candidates ($B^0$ or
$B^+$) or $\Bbar$ candidates ($\Bbar^0$ or $B^-$), except for ambiguous
modes with a $\KS$ and zero net charge.  In order to correct the
imperfect knowledge of the hadronic final state ingredients, the signal
yield for each exclusive mode is used to correct the Monte Carlo
multiplicity distribution.  The resulting $\Bbar$-tagged
($342\pm23\PM{7}{14}$ events), $B$-tagged ($349\pm23\PM{7}{14}$ events)
and ambiguous ($47.8\pm8.7\PM{1.4}{1.8}$ events) signals are shown in
Fig.~\ref{fig:belle-acpxsgam}.  Using the wrong-tag fractions of
$0.019\pm0.014$ between $B$- and $\Bbar$-tagged, $0.240\pm0.192$ from
ambiguous to $B$- or $\Bbar$-tagged, and $0.0075\pm0.0079$ from $B$- or
$\Bbar$-tagged to ambiguous samples, the asymmetry is measured to be
\begin{equation}
\Acp(\BtoXsgamma)=0.004\pm0.051\pm0.038.
\end{equation}
The result corresponds to a 90\% confidence level limit of
$-0.107<\Acp(\BtoXsgamma)<0.099$, and therefore already constrains
extreme cases of the new physics parameter space.

For exclusive radiative decays, it is straightforward to extend the
analysis to search for direct CP-asymmetry\rlap.\,\cite{bib:kstargam-cleo}$^-$\cite{bib:kstargam-belle}
Particle identification devices of Belle and BaBar resolve the possible
ambiguity between $K^{*0}\to\KP\piM$ and $\Kbar^{*0}\to\KM\piP$ to an
almost negligible level with a reliable estimation of the wrong-tag
fraction (0.9\% for Belle).  The results of the asymmetry measurements are
listed in Table~\ref{tbl:acpkstargam}, whose average is
\begin{equation}
\Acp(\BtoKstargamma)=(-0.5\pm3.7)\times10^{-2}.
\end{equation}
It is usually considered that the large CP-violation in $\BtoKstargamma$
is not allowed in the SM and the result may be used to constrain new
physics.  However, as the strong phase difference involved may not be
reliably calculated for exclusive decays, the interpretation may be model
dependent.

\begin{table}
\begin{center}
\caption{$B\to K^*\gamma$ direct CP-asymmetry}
\label{tbl:acpkstargam}
\begin{tabular}{cc}
\hline
CLEO  $(9.1\fbinv)$  & $(8\pm13\pm3)\times10^{-2}$ \\
BaBar $(20.7\fbinv)$ & $(-4.4\pm7.6\pm1.2)\times10^{-2}$ \\
Belle $(78\fbinv)$   & $(-0.1\pm4.4\pm0.8)\times10^{-2}$ \\
\hline
\end{tabular}
\end{center}
\end{table}


\subsection{Search for $\btodgamma$ Final States}

There are various interesting aspects in the $\btodgamma$ transition.
Within the SM, most of the diagrams are a copy of those for
$\btosgamma$, except for the replacement of the CKM matrix element
$\Vts$ with $\Vtd$.  A measurement of the $\btodgamma$ process will
therefore provide the ratio $|\Vtd/\Vts|$ without large model dependent
uncertainties.  This is in contrast with the current best $|\Vtd/\Vts|$
limit obtained from $\Delta m_s$ and $\Delta m_d$ in $B_s$ and $B_d$
mixing with the help of lattice QCD calculations.  Unfortunately,
the inclusive $\BtoXdgamma$ measurement is extremely difficult due to its
small rate and the huge $\BtoXsgamma$ background, and the use of
exclusive decay modes such as $B\to\rho\gamma$ involves other model
dependences.  If the constraints of the SM is relaxed, it is not
necessary to retain the CKM structure, and $\btodgamma$ becomes a
completely new probe to search for new physics effects in the $b\to d$
transition that might be hidden in the $B_d$ mixing and cannot be
accessed in the $b\to s$ transition.  This mode is also the place where
a large direct CP-asymmetry is predicted within and beyond the SM.

The search for the exclusive decay $B\to\rho\gamma$ is as straightforward as
the measurement of $\BtoKstargamma$, except for its small branching
fraction, the enormous combinatorial background from copious $\rho$
mesons and random pions, and the huge $\BtoKstargamma$ background that
overlaps with the $B\to\rho\gamma$ signal window.  BaBar has optimized
the background suppression algorithm using a neural net technique with
input parameters of the event shape, helicity angle and vertex
displacement between the signal candidate and the rest of the event, and
has optimized the kaon rejection algorithm so that the $\BtoKstargamma$
background can be suppressed to a negligible level.  $B\to\omega\gamma$
is not affected by $\BtoKstargamma$, but it is still hardly observed.
The upper limits obtained by BaBar\CITE,{bib:rhogam-babar}
Belle\cite{bib:rhogam-belle} and CLEO\cite{bib:kstargam-cleo} are
summarized in Table~\ref{tbl:rhogam}.  The best upper limits by BaBar
are still about twice as large as the SM
predictions\CITE,{bib:kstargam-sm} $(9.0\pm3.4)\times10^{-7}$ for
$\rho^+\gamma$, and $(4.9\pm1.8)\times10^{-7}$ for $\rho^0\gamma$ and
$\omega\gamma$.

\begin{table}
\begin{center}
\caption{90\% confidence level upper limits on the $B\to \rho\gamma$ and
$\omega\gamma$ branching fractions.}
\label{tbl:rhogam}
\begin{tabular}{cccc}
\hline
& $\rho^+\gamma$ & $\rho^0\gamma$ & $\omega\gamma$ \\
\hline
CLEO  $(9.1\fbinv)$ & $13\EM6$  & $17\EM6$  & $9.2\EM6$ \\
Belle $(78\fbinv)$  & $2.7\EM6$ & $2.6\EM6$ & $4.4\EM6$ \\
BaBar $(78\fbinv)$  & $2.1\EM6$ & $1.2\EM6$ & $1.0\EM6$ \\ 
\hline
\end{tabular}
\end{center}
\end{table}

\begin{figure}
  \center
  \psfig{figure=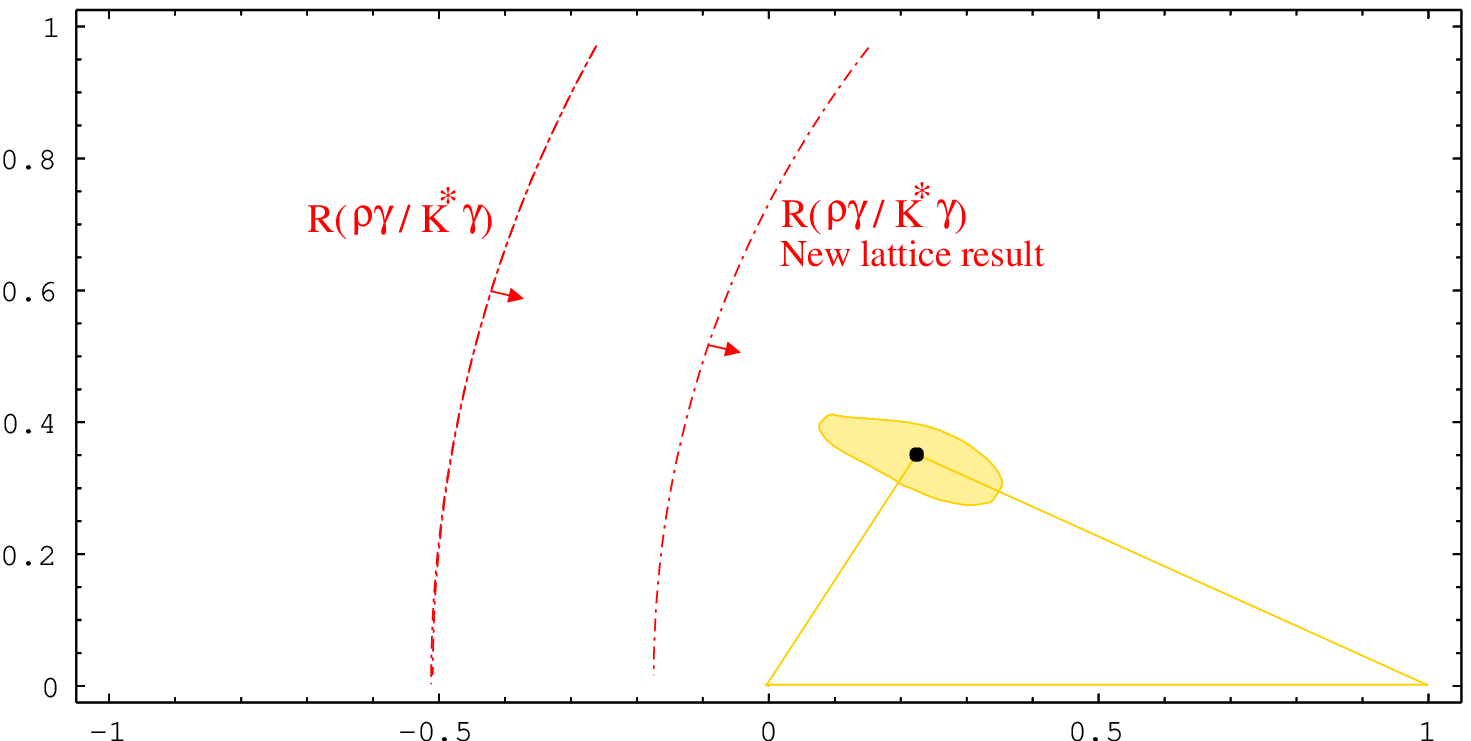,width=8truecm}
  \caption{Bound on $\Vtd$ from the ratio of branching fractions of
  $\Btorhogamma$ to $\BtoKstargamma$.}
  \label{fig:rhogam-vtd}
\end{figure}

Using the isospin relation
$\Gamma(\Btorhogamma)\equiv\Gamma(\Btorhoplusgamma)=2\Gamma(\Btorhozerogamma)$,
the combined $\Btorhogamma$ upper limit from BaBar becomes
$\Br(\Btorhogamma)<1.9\EM6$.  The ratio of the branching fractions can
be expressed as
\begin{equation}
\begin{array}{rcl}\displaystyle
{\Br(\Btorhogamma)\over\Br(\BtoKstargamma)}
&=&\displaystyle
   \left|{\Vtd\over\Vts}\right|^2
   \left({m_B^2-m_\rho^2\over m_B^2-m_{K^*}^2}\right)^3
   \zeta^2 [ 1 + \Delta R ]\\[18pt]
&<&0.047 \mbox{~(90\%~CL)}
\end{array}
\end{equation}
where $\zeta=0.76\pm0.10$ is the ratio of the form factors obtained from
the light-cone QCD sum rule and $\Delta R=0.0\pm0.2$ is to account for
$SU(3)$ breaking effects.  From this inequality, a bound on $\Vtd$ is
given as $|\Vtd/\Vts|<0.34$, which is still a weaker constraint than
that given by $\Delta m_s/\Delta m_d$.  One can still argue about the
validity of the form factor ratio\CITE,{bib:rhogam-lunghi} as a recent
lattice QCD calculation\cite{bib:kstargam-becirevic} gives a value of
$\zeta=0.91\pm0.08$ that leads to a different constraint on $\Vtd$ as
shown in Fig.~\ref{fig:rhogam-vtd}.

\section{Electroweak Rare $B$ Decays}

The $\btosll$ transition has a lepton pair in the final state, which is
a clear signature of the decay.  The decay amplitude is written down
using three Wilson coefficients, $C_7$, $C_9$ and $C_{10}$.  Although
there are three unknown complex coefficients ($|C_7|$ is obtained from
$\BtoXsgamma$), it is possible to disentangle all of them from
measurements of the $\shat=q^2/m_b^2$ dependent branching fraction
$d\Gamma/d\shat$ and the forward-backward asymmetry $d\AFB/d\shat$,
\begin{equation}
\begin{array}{rl}
\displaystyle
{d\Gamma\over d\shat}
&\displaystyle
  = \left(\alpha_{\rm em}\over 4\pi\right)^2
   {G_F^2 m_b^5 \left|V_{ts}^*V_{tb}\right|^2\over 48\pi^3}
   (1-\shat)^2
\\[6pt]
\displaystyle
   &
   \times
   \left[ (1+2\shat)\left(|\Cnine|^2 + |\Cten|^2\right)
       + 4\left(1+{2\over\shat}\right)|\Cseven|^2 \right.\\
\displaystyle &
   \left. + 12{\rm Re}\left(\Cseven\Cnine\right) \right ] + \mbox{corr.}\\
\end{array}
\end{equation}
\begin{equation}
{d\AFB\over d\shat}=\Cten(2\Cseven+\Cnine\shat) / (d\Gamma / d\shat),
\end{equation}
where QCD corrections are included in the Wilson coefficients.

There are two amplitudes that interfere with $\btosll$: one is
$\btosgamma$ at $q^2\to0$ and the other is $b\to(\ccbar)s$ where
$(\ccbar)$ is a charmonium state such as $J/\psi$ or $\psi'$ that decays
into $\elel$.  The latest theory calculation that includes
Next-to-Next-to-Leading-Order QCD corrections has been completed for the
restricted range of $0.05<\shat<0.25$ to avoid these
interferences\CITE.{bib:xsll-sm}

Similarly to $\btosgamma$, there are a number of extensions to the
SM\cite{bib:xsll-bsm} that one may be sensitive to by studying $\btosll$, and
$B$-factories have just opened the window to search for such effects
with a huge sample of $B$ decays that was not available before.


\subsection{Observation of $\BtoKstarll$}

The first signal of $\BtoKll$ was observed by Belle\cite{bib:kll-belle}
using $29\fbinv$ data and confirmed by BaBar\cite{bib:kll-babar} with
$78\fbinv$, while the $\BtoKstarll$ signal, which has a larger expected branching
fraction, was not significant with those data samples.

The $\BtoKorKstarll$ signal is identified with $\Mbc$, $\DeltaE$ (and
$\MKpi$ for $K^*\elel$).  There are five types of background that may
contribute.  1) Charmonium decays, $B\to J/\psi K^{(*)}$ and $\psi'
K^{(*)}$ have to be removed by the corresponding $\Mll$ veto windows
around $J/\psi$ and $\psi'$ masses.  Especially for $\epem$ modes,
bremsstrahlung has to be taken into account.  2) Hadronic decays, $B\to
K^{(*)}\piP\piM$, are almost completely removed by lepton selection criteria
including minimum lepton momentum requirements, but the remaining small
contribution has to be evaluated and subtracted from the signal peak.
3) Two leptons from semi-leptonic decays, either in the cascade $b\to
c\to s,d$ chain or from two $B$ mesons, combined with a random $K^{*}$.
This is the dominant combinatorial background that can be reduced for
example by using the missing energy of the event.  4) Continuum
background, which can be reduced by shape variables.  5) Rare
backgrounds, $K^*\gamma$ with a photon conversion to $\epem$, and
$K^{(*)}\piZ$ with a $\pi^0$ decaying into $\epem\gamma$.  This
background can be removed by requiring a minimum $\epem$ mass as is done by
Belle or can be subtracted from the signal as done by BaBar.

\begin{figure}
  \center
  \psfig{figure=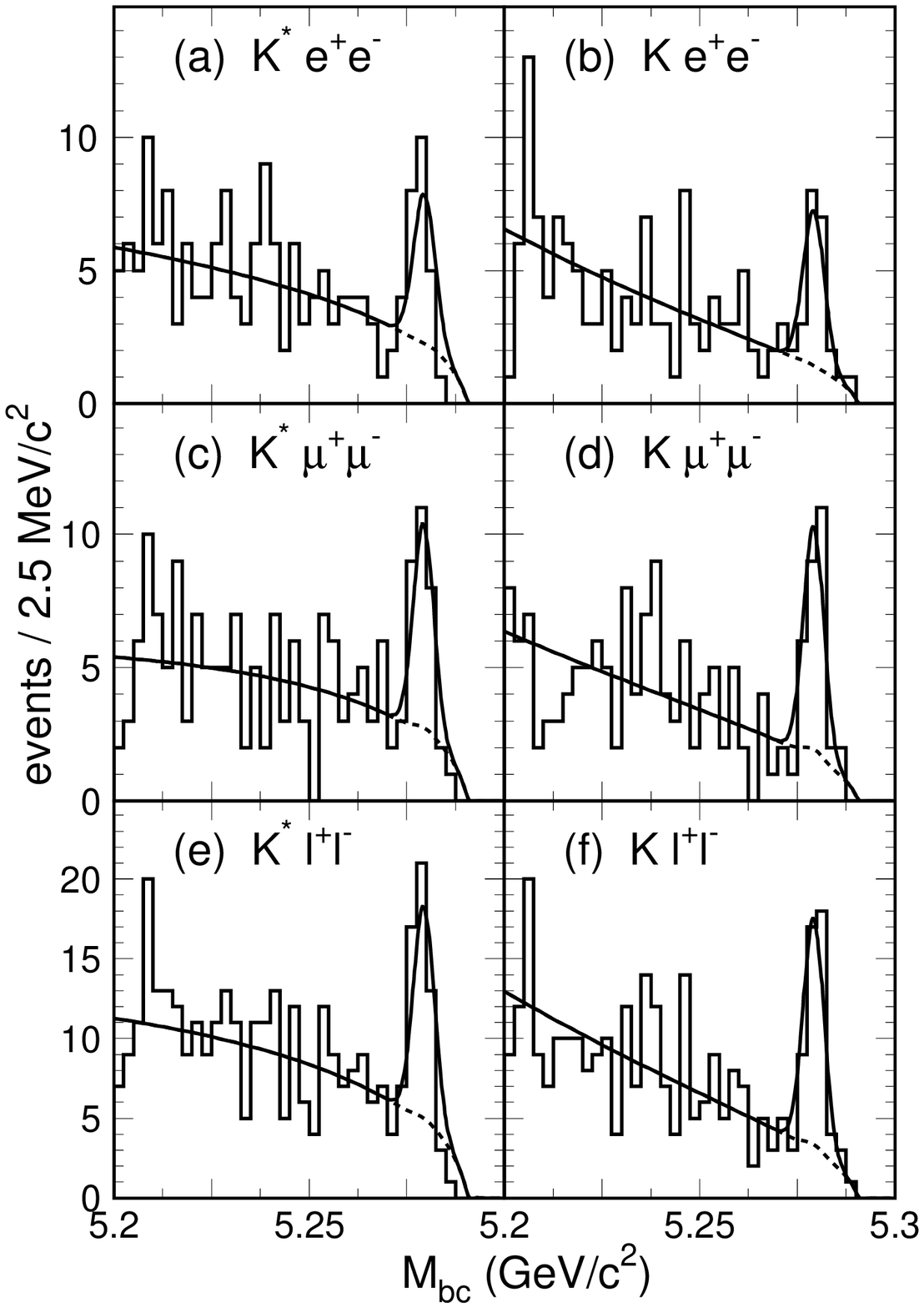,width=7truecm}
  \caption{$\BtoKorKstarll$ signal observed by Belle.}
  \label{fig:kstarll-belle}
\end{figure}

Belle has updated the analysis using a $140\fbinv$ data sample, with a
number of improvements in the analysis
procedure\CITE.{bib:kstarll-belle}  The most significant improvement is
the lowered minimum lepton momentum of 0.7 (0.4) GeV for muons (electrons)
from 1.0 (0.5) GeV to gain 12\% (7\%) in the total efficiency.  In
addition, a $K^*\elel$ combination is removed if there can be an
unobserved photon along with one of the leptons that can form $B\to J/\psi
K\to \elel\gamma K$.  As a result, the first $\BtoKstarll$ signal is
observed with a statistical significance of 5.7 from a fit to $\Mbc$, as
shown in Fig.~\ref{fig:kstarll-belle}, together with the improved
$\BtoKll$ signal with a significance of 7.4.

\begin{figure}
  \center
  \psfig{figure=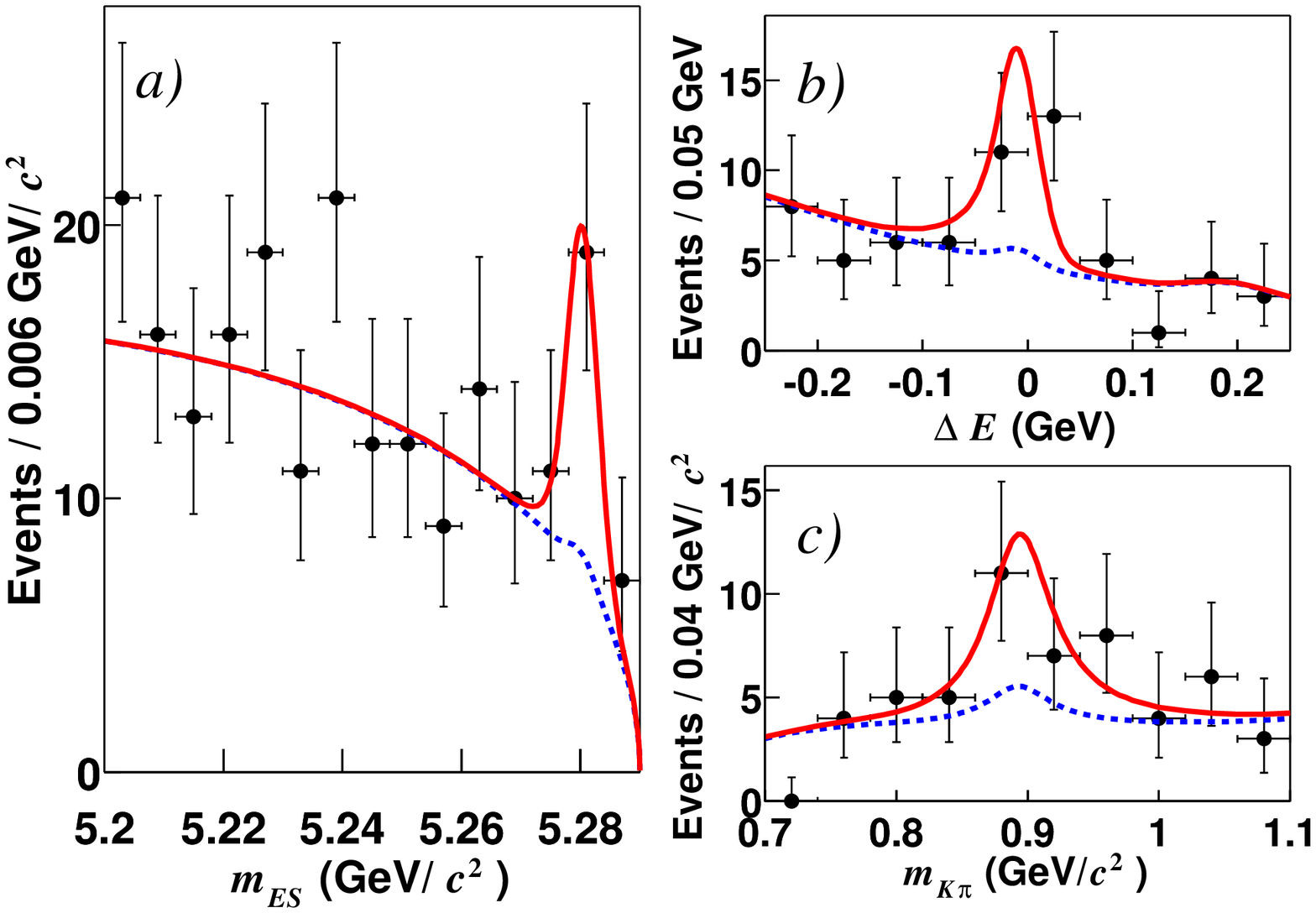,width=8truecm}
  \caption{Evidence for $\BtoKstarll$ from BaBar.}
  \label{fig:kstarll-babar}
\end{figure}

\begin{figure}
  \center
  \psfig{figure=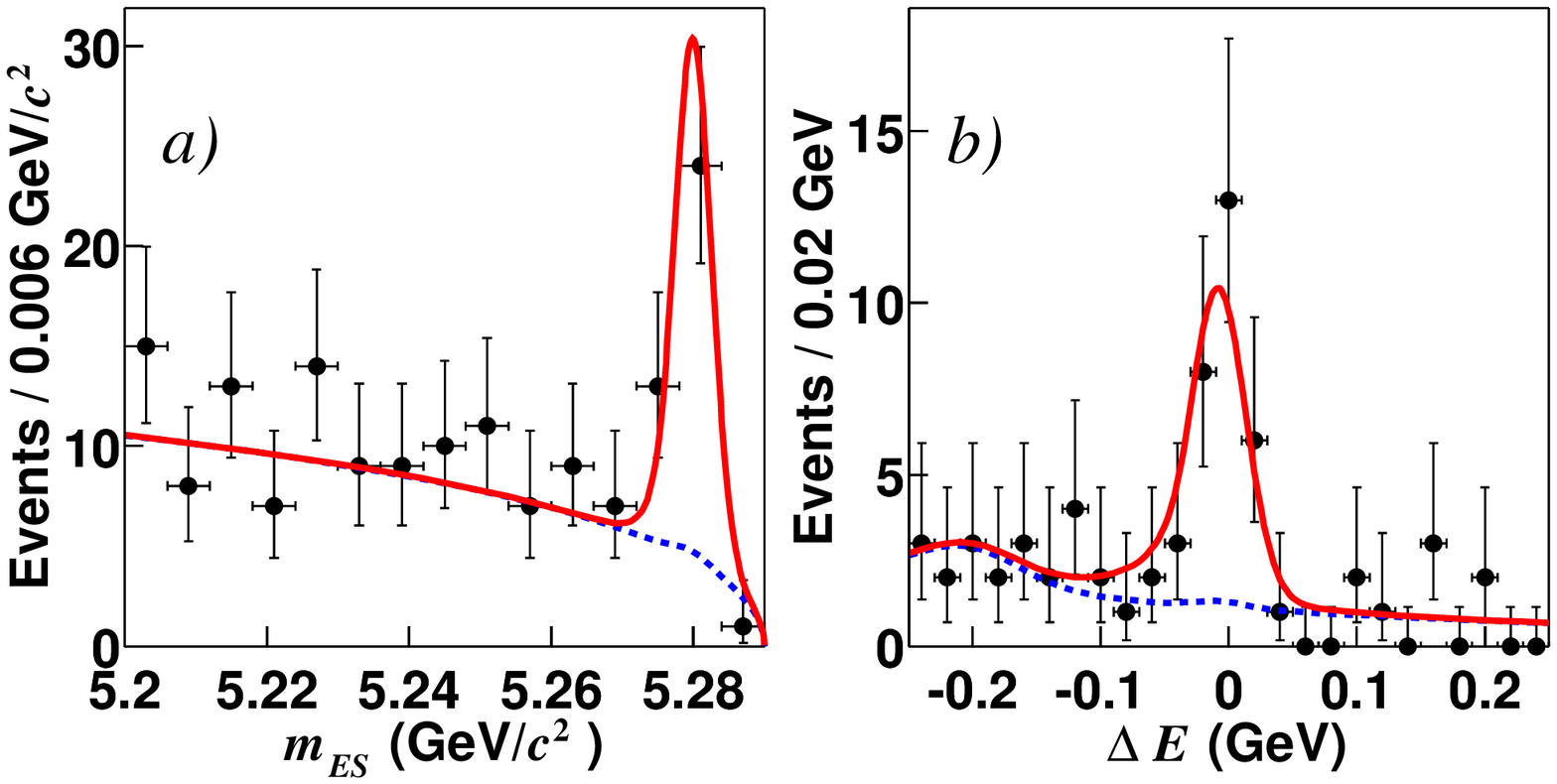,width=8truecm}
  \caption{Signal for $\BtoKll$ from BaBar.}
  \label{fig:kll-babar}
\end{figure}

BaBar has also updated the analysis using a $113\fbinv$ data
sample\CITE,{bib:kstarll-babar} with improvements such as the
bremsstrahlung photon recovery to include $K^{(*)}\epem\gamma$ events in
the $K^{(*)}\epem$ signal.  Evidence for the $\BtoKstarll$ signal is
also seen with a statistical significance of 3.3 from a simultaneous fit
to $\Mbc$, $\DeltaE$ and $\MKpi$ (Fig.~\ref{fig:kstarll-babar} shows
their projections).  A signal for $\BtoKll$ is clearly observed with a
significance of ${\sim}8$ (Fig.~\ref{fig:kll-babar}).

The branching fractions obtained are summarized in
Table~\ref{tbl:kstarll}.  For the combined $\BtoKstarll$ results,
$\Br(\BtoKstarll)=\Br(\BtoKstarmumu)=0.75\Br(\BtoKstaree)$ is assumed which
compensates the enhancement at the $q^2=0$ pole that appears more
significantly in $K^*\epem$, using the expected SM
ratio\CITE.{bib:kstarll-ali}  The measured branching fractions are in
agreement with the SM, for example\cite{bib:kstarll-ali}
$(3.5\pm1.2)\EM7$ for $\BtoKll$ and $(11.9\pm3.9)\EM7$ for
$\BtoKstarll$.  We note that the experimental errors are already much
smaller than the uncertainties in the SM predictions\cite{bib:kll-sm}
and the variations due to different model-dependent assumptions used to
account for the hadronic uncertainties.

\begin{table}
\caption{$\BtoKorKstarll$ branching fractions.}
\label{tbl:kstarll}
\center
\begin{tabular}{lcc}
\hline
Mode & Belle ($140\fbinv$) & BaBar ($113\fbinv$) \\
     & $[\EM7]$ & $[\EM7]$ \\
\hline
  & \\[-0.30cm]
$B\to K\epem$
  & $4.8^{+1.5}_{-1.3}\pm0.3\pm0.1$
  & $7.9^{+1.9}_{-1.7}\pm0.7$ \\[0.05cm]
$B\to K\mumu$
  & $4.8^{+1.3}_{-1.1}\pm0.3\pm0.2$
  & $4.8^{+2.5}_{-2.0}\pm0.4$ \\[0.05cm]
$B\to K\elel$
  & $4.8^{+1.0}_{-0.9}\pm0.3\pm0.1$
  & $6.9^{+1.5}_{-1.3}\pm0.6$ \\[0.05cm]
\hline
  & \\[-0.30cm]
$B\to K^*\epem$
  & $14.9^{+5.2+1.1}_{-4.6-1.3}\pm0.3$
  & $10.0^{+5.0}_{-4.2}\pm1.3$ \\[0.05cm]
$B\to K^*\mumu$
  & $11.7^{+3.6}_{-3.1}\pm0.8\pm0.6$
  & $12.8^{+7.8}_{-6.2}\pm1.7$ \\[0.05cm]
$B\to K^*\elel$
  & $11.5^{+2.6}_{-2.4}\pm0.7\pm0.4$
  & $8.9^{+3.4}_{-2.9}\pm1.1$ \\[0.05cm]
\hline
\end{tabular}
\end{table}

\begin{figure}
  \center
  \psfig{figure=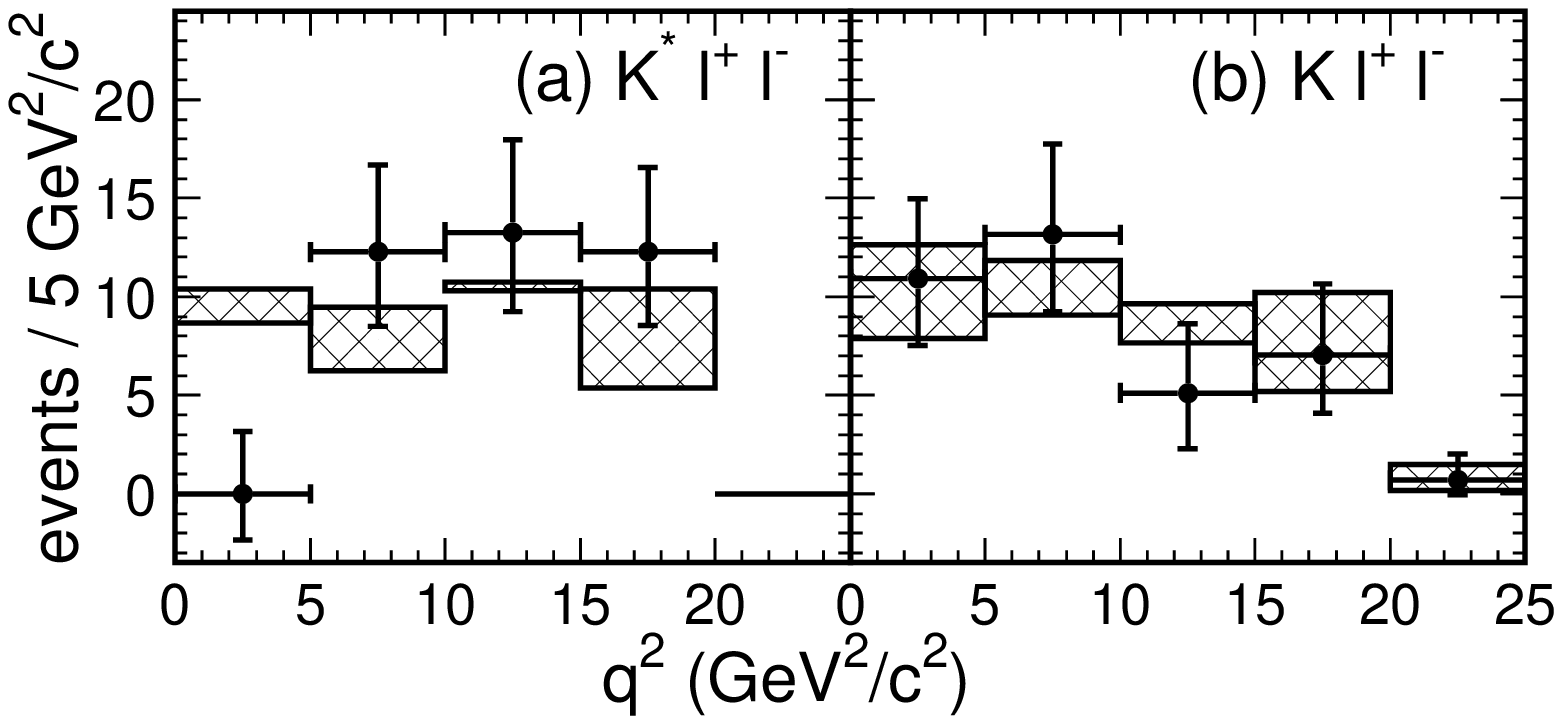,width=8truecm}
  \caption{$q^2$ distributions for $\BtoKorKstarll$ from Belle.}
  \label{fig:belle-kll-q2}
\end{figure}

It is still too early to fit the $q^2$ distribution to constrain new
physics.  First attempts to extract the $q^2$ distribution using the
individual $\Mbc$ signal yields in $q^2$ bins has been performed by Belle as
shown in Fig.~\ref{fig:belle-kll-q2}.

\subsection{Measurement of $\BtoXsll$}

The first measurements of the $\BtoKorKstarll$ branching fractions are
consistent with the SM predictions. However since these predictions have uncertainties that are already
larger than the measurement errors, the inclusive rate for $\BtoXsll$
becomes more important in terms of the search for a deviation from the SM.
In contrast to $\BtoXsgamma$, the lepton pair alone does not provide a
sufficient constraint to suppress the largest background from
semi-leptonic decays.  Therefore, it is only possible to use the
semi-inclusive method to sum up the exclusive modes for now.

Belle has successfully measured the inclusive $\BtoXsll$ branching
fraction\cite{bib:xsll-belle} from a $60\fbinv$ data sample by
applying the method to sum up the $X_s$ final state with one kaon ($K^+$
or $\KS$) and up to four pions, of which one pion is allowed to be
$\pi^0$.  Assuming the $\KL$ contribution is the same as $\KS$, this set
of final states covers $82\pm2\%$ of the signal.  In addition, $\MXs$ is
required to be below $2.1\GeV$ in order to reduce backgrounds.  For
leptons, minimum momentum of $0.5\GeV$ for electrons, $1.0\GeV$ for
muons and $M(\elel)>0.2\GeV$ are required.  Background sources and the
suppression techniques are similar to the exclusive decays.

A new result is reported by BaBar with a $78\fbinv$ data sample, using
the same method with slightly different
conditions\CITE.{bib:xsll-babar}  BaBar includes up to two pions,
corresponding to 75\% of the signal, and require $\MXs<1.8\GeV$.
The minimum muon momentum requirement of $0.8\GeV$ is lower than Belle's.

\begin{figure}[th]
  \center
  \psfig{figure=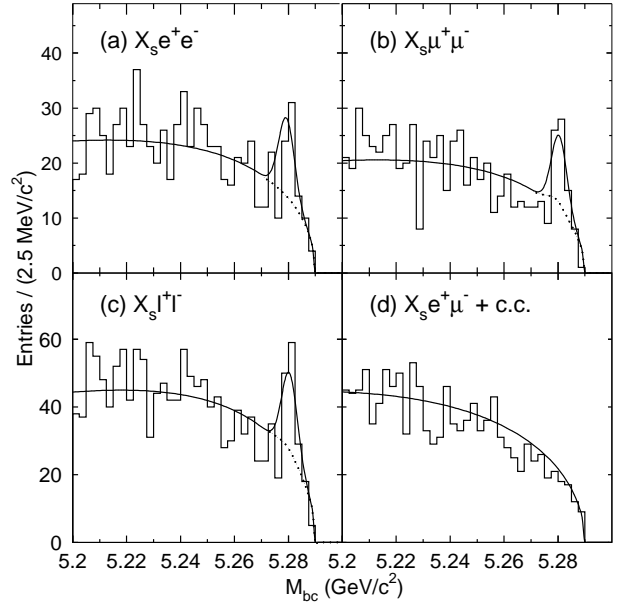,width=8truecm}
  \caption{$\BtoXsll$ signal measured from Belle.  The $X_s e^+\mu^-$ sample,
  which is prohibited in the SM, represents the combinatorial backgrounds.}
  \label{fig:xsll-belle}
\end{figure}

\begin{figure}[th]
  \center
  \psfig{figure=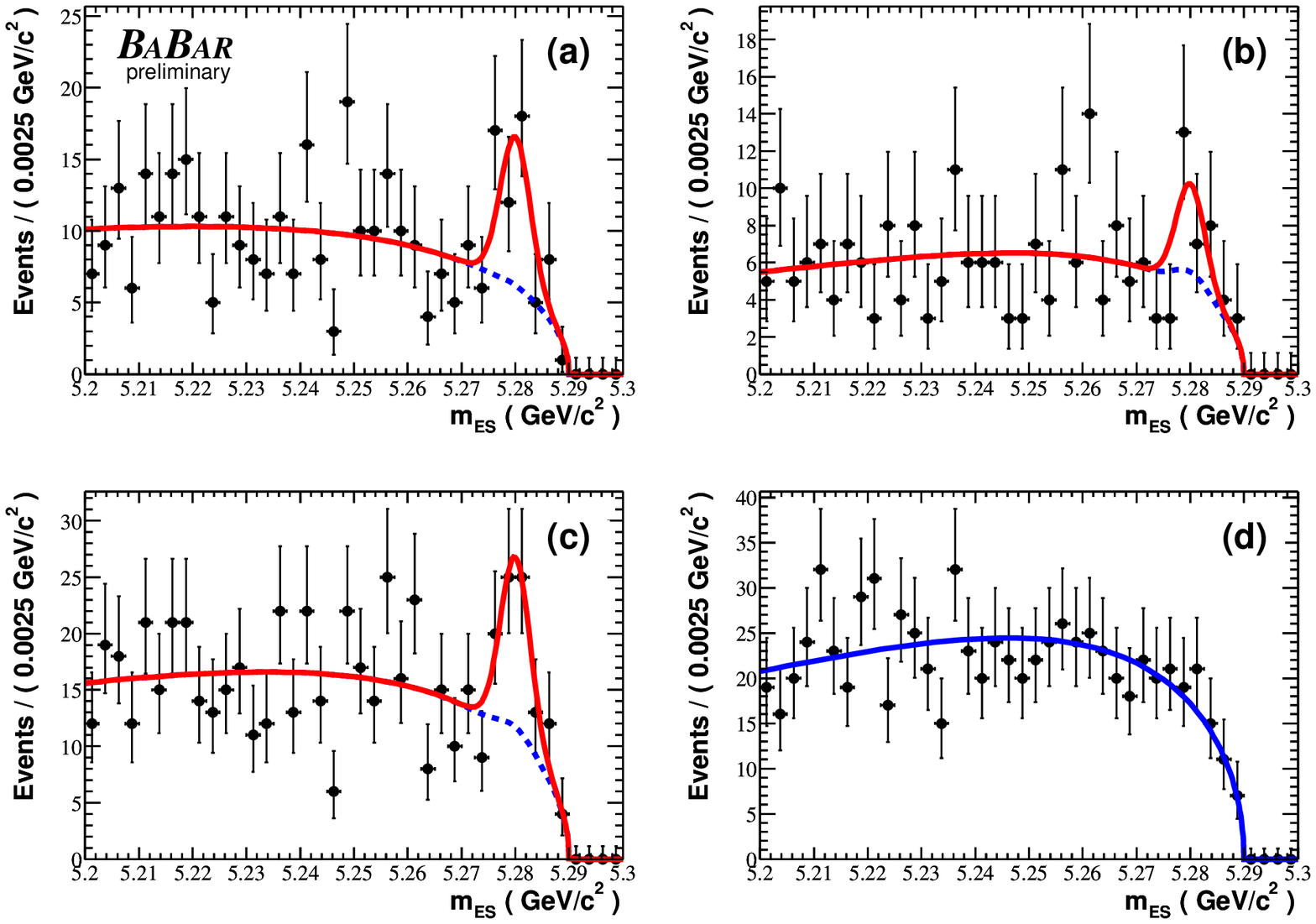,width=8truecm}
  \caption{(a) $\BtoXsee$, (b) $\BtoXsmumu$, (c) $\BtoXsll$ signals with 
  the (d) $X_s e^+\mu^-$ sample from BaBar.}
  \label{fig:xsll-babar}
\end{figure}

The signal of $60\pm14$ events from Belle with a statistical significance
of 5.4 is shown in Fig.~\ref{fig:xsll-belle}, and $41\pm10$ events from
BaBar with a significance of 4.6 is shown in Fig.~\ref{fig:xsll-babar}.
Corresponding branching fractions are very close to each other as given in
Table~\ref{tbl:xsll}, whose average is
\begin{equation}
\Br(\BtoXsll)=(6.2\pm1.1\PM{1.6}{1.3})\EM6.
\end{equation}
The branching fraction results are for the dilepton mass range above
$\Mll>0.2\GeV$ and are interpolated in the $J/\psi$ and $\psi'$ regions
that are removed from the analysis, assuming no interference with these
charmonium states.

\begin{table}
\caption{$\BtoXsll$ branching fractions.}
\label{tbl:xsll}
\begin{center}
\begin{tabular}{lcc}
\hline
Mode & Belle ($60\fbinv$) & BaBar ($78\fbinv$) \\
     & $[\EM6]$ & $[\EM6]$ \\
\hline
$\Xsee$   & $5.0\pm2.3\PM{1.3}{1.1}$ & $6.6\pm1.9\PM{1.9}{1.6}$ \\
$\Xsmumu$ & $7.9\pm2.1\PM{2.1}{1.5}$ & $5.7\pm2.8\PM{1.7}{1.4}$ \\
$\Xsll$   & $6.1\pm1.4\PM{1.4}{1.1}$ & $6.3\pm1.6\PM{1.8}{1.5}$ \\
\hline
\end{tabular}
\end{center}
\end{table}

\begin{figure}
  \center
  \psfig{figure=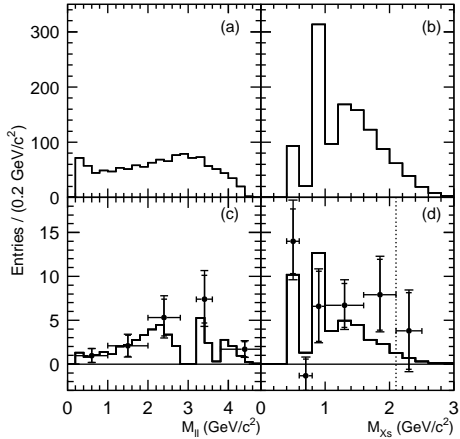,width=6truecm}
  \caption{$\Mll$ (left) and $\MXs$ (right) distributions for $\BtoXsll$
  from Belle (points with error bars), compared with the SM predictions
  before (top) and after (bottom) including detector acceptance effects.}
  \label{fig:xsll-dist-belle}
\end{figure}

\begin{figure}
  \center
  \psfig{figure=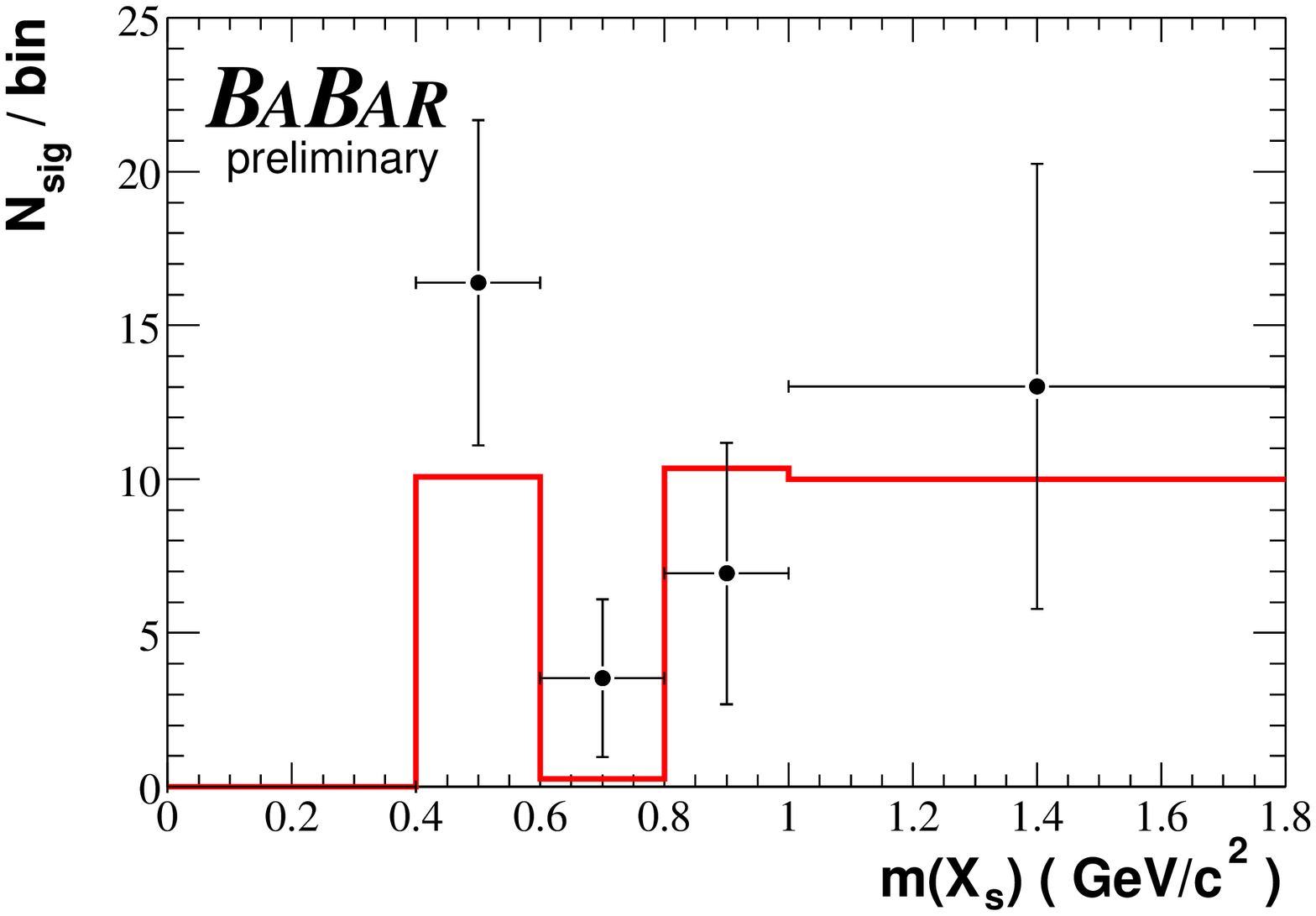,width=6truecm}
  \caption{$\MXs$ distribution for $\BtoXsll$ from BaBar.}
  \label{fig:xsll-dist-babar}
\end{figure}

The results may be compared with the SM prediction\cite{bib:xsll-sm-ali}
of $(4.2\pm0.7)\EM6$ integrated over the same dilepton mass range of
$\Mll>0.2\GeV$.  With this requirement, the effect of the $q^2=0$ pole
becomes insignificant, giving almost equal branching fractions for the
electron and muon modes.  The measured branching fractions are in agreement
with the SM, considering the large measurement error.  It should be
noted that the large systematic error is dominated by the uncertainty in
the $\MXs$ distribution, in particular the fraction of $\BtoKorKstarll$,
that will be reduced with more statistics.  Distributions for $\MXs$ and
$\Mll$ are shown in Figs.~\ref{fig:xsll-dist-belle} and
\ref{fig:xsll-dist-babar}, in which no significant deviation from the SM
is observed.

\subsection{Search for $B\to K\nu\nubar$}

The $b\to s\nu\nubar$ channel is sensitive to the weak-boson part of the
$\btosll$ amplitude, and does not involve the $q^2=0$ pole and
interfering charmonium decays.  It is experimentally challenging even
for the easiest exclusive $B^+\to K^+\nu\nubar$ channel, because there
is only one measurable kaon track out of the three-body final state
that characterize the signal.  In order to identify the signal, the
other side $B$ decay has to be tagged, so that there is only one kaon in
the rest of the event.  The search is attempted by BaBar using two
techniques to tag the other $B$.

\begin{figure}
  \center
  \psfig{figure=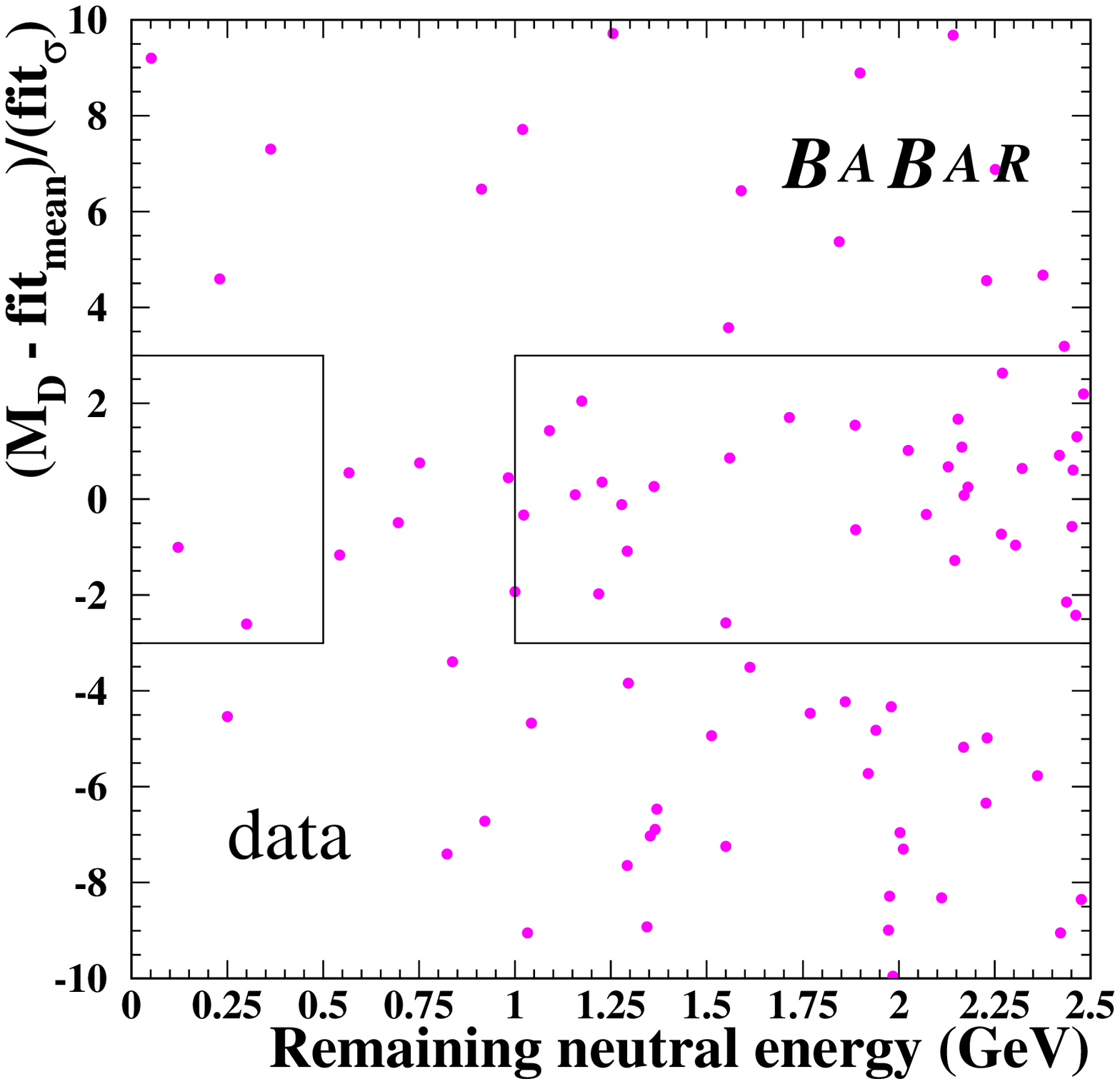,width=5truecm}
  \caption{$B\to K\nu\nubar$ search results from BaBar using the
  semi-leptonic tag technique.}
  \label{fig:knunu}
\end{figure}

One method\cite{bib:knunu-semilep} is to require a $D^0$ meson and a
lepton in the event that come from the $B^-\to D^{(*)0}\ell^-\nubar$
decay channel to tag the semi-leptonic decay of the other side $B$.
After removing the signal kaon and the tag-side $D^0$ and lepton, there
should be no remaining charged tracks, and the energy in the calorimeter
should be at most that from the disregarded soft photon or $\piZ$ from
the $D^{*0}$ decay.  The signal window is defined in the plane of the
remaining energy (less than $0.5\GeV$) and the reconstructed $D^0$ mass
(within $\pm3\sigma$).  Two candidates are found using a $51\fbinv$ data
sample (Fig.~\ref{fig:knunu}) with a tagging efficiency of 0.5\%, where
2.2 background events are expected.  This leads to the upper limit of
$\Br(B\to K\nu\nubar)<9.4\EM5$ at the 90\% confidence level.

The other method\cite{bib:knunu-hadronic} is to require a full
reconstruction of the hadronic decay $B^-\to D^0 X^-$, where $X^-$
represents a combination of up to three charged pions or kaons and up to
two $\piZ$ with a net charge of $-1$ to tag the hadronic decay of the
other side $B$.  In this case, the maximum remaining energy is reduced
to $0.3\GeV$.  The signal is identified with a high energy kaon with
more than $1.5\GeV$.  Three candidates are found using a $80\fbinv$ data
sample with a tagging efficiency of 0.13\%, where $2.7\pm0.8$ background
events are expected.  This leads to the upper limit of $\Br(B\to
K\nu\nubar)<10.5\EM5$ at the 90\% confidence level.

Since the two methods use statistically independent sub-samples, the two
results can be combined to improve the upper limit as
\begin{equation}
\Br(B\to K\nu\nubar)<7.0\EM5\mbox{~(90\% C.L.)},
\end{equation}
which is still an order of magnitude higher than the SM
prediction\cite{bib:knunu-sm} of $\Br(B\to
K\nu\nubar)=(3.8\PM{1.2}{0.6})\EM6$.


\section{Pure Leptonic $B$ Decays}

Leptonic two-body $B$ decays are highly helicity suppressed in the SM
due to the large energy release from the $B$ meson decaying into much
lighter leptons.  The branching fraction for $B^+\to\ell^+\nu$ is
written down as
\begin{equation}
\Br(B^+\to\ell^+\nu)
 = {G_F^2 m_B\over8\pi}m_l^2\left(1-{m_l^2\over m_B^2}\right)^2
   f_B|\Vub|^2\tau_B
\end{equation}
which is sensitive to $\Vub$ and the $B$ meson decay constant $f_B$.
The experimental sensitivities are still far above the predicted SM
branching fractions.

However, if there is a non-SM decay amplitude that is not helicity
suppressed, the branching fraction may be accessible by the on-going
experiments.  The decay modes considered are: $B^+\to\tau^+\nu$,
$B^+\to\mu^+\nu$, $B_d^0\to\mu^+\mu^-$, $B_d^0\to e^+e^-$ and
$B_s^0\to\mu^+\mu^-$.  The lepton flavor violating $B_d^0\to e^\pm\mu^\mp$
is also searched for.


\subsection{Search for $B\to\tau\nu$ and $B\to\mu\nu$}

The decay $B^+\to\tau^+\nu$ has been searched for by BaBar using
$81\fbinv$ of data.  As there are at least two missing neutrinos, the same
two techniques for the $B\to K\nu\nubar$ search are applied to tag the
other side $B$ using semi-leptonic decays and hadronic decays.

\begin{figure}
  \center
  \psfig{figure=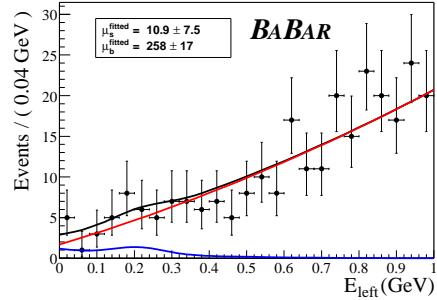,width=6truecm}
  \caption{$B^+\to \tau^+\nu$ search results from BaBar using the
  semi-leptonic tag technique.}
  \label{fig:taunu}
\end{figure}

In the analysis with the leptonic tag, $\tau^+\to e^+\nu_e\nubar_\tau$
and $\mu^+\nu_\mu\nubar_\tau$ are used\CITE.{bib:taunu-semilep}  A fit to
the remaining energy, that can include a soft $\gamma/\piZ$ in the other
side $B$, shows no significant excess above the expected background
(Fig.~\ref{fig:taunu}).  The upper limit is obtained to be
$\Br(B^+\to\tau^+\nu)<7.7\EM4$ at the 90\% confidence level.

In the analysis with the hadronic tag, hadronic $\tau$ decays into
$\pi^+\nubar_\tau$, $\pi^+\pi^0\nubar_\tau$ and
$\pi^+\pi^-\pi^+\nubar_\tau$ are also included\CITE.{bib:taunu-hadronic}
The number of events with the remaining energy less than $\sim100\MeV$
is counted.  In total 35 candidates are found for the expected
background of $37.6\pm4.7\pm1.3$ events.  The upper limit is obtained to
be $\Br(B^+\to\tau^+\nu)<4.9\EM4$ at the 90\% confidence level.

These two samples are combined to improve the upper limit,
\begin{equation}
\Br(B^+\to \tau^+\nu_\tau)<4.1\EM4\mbox{~(90\%~CL)}.
\end{equation}
This improves the previous upper limit given by L3\CITE.{bib:taunu-l3}
The corresponding SM prediction is $7.5\EM5$ for $\tau_B=1.674\mbox{~ps}$,
$f_B=198\MeV$ and $|\Vub|=0.0036$.

The decay $B^+\to\mu^+\nu$ has been searched for by
Belle\cite{bib:munu-belle} and BaBar\CITE.{bib:munu-babar}
The analysis technique is to use the ``neutrino reconstruction''
technique, to determine the neutrino momentum from the missing momentum
of the event.  The muon momentum is monochromatic, except for the small
initial $B$ meson momentum of $340\MeV$, while the muon from the
dominant background source of semi-leptonic $B$ decays have a smaller
momentum.  No significant signal excess has been observed; the most
stringent upper limit is given by BaBar using $81\fbinv$,
\begin{equation}
\Br(B^+\to \tau^+\nu_\tau)<6.6\EM6\mbox{~(90\%~CL)}.
\end{equation}
The SM predicts an order of magnitude smaller branching fraction of
$\sim4\EM7$.


\subsection{Search for $B\to\elel$}

In the SM, the decay $B_{d,s}^0\to\elel$ occurs through the electroweak
penguin transition $b\to(d,s)\elel$, and due to the helicity
suppression, the expected branching fraction is extremely
small:\cite{bib:ll-sm} $(2.34\pm0.33)\EM{15}$ for $B_d^0\to\epem$,
$(1.00\pm0.14)\EM{10}$ for $B_d^0\to\mumu$ and $(3.4\pm0.5)\EM9$ for
$B_s^0\to\mumu$. The decay amplitude may be significantly enhanced in
some extensions to the SM.  For example, these decays are sensitive to
the chirality flipping interaction in models with two Higgs doublets,
and the branching fractions can be three orders of magnitude larger than
the SM at large $\tan\beta$, and may be accessible by the $B$-factories
for $B_d^0$ decays and by the Tevatron for $B_s^0$ decays.  In this case the
$\BtoXsll$ decay rate may not be affected and can be consistent with the
SM.  The search can be easily extended to the lepton flavor violating
decay $B_d^0\to e^\pm\mu^\mp$.

\begin{figure}
  \center
  \psfig{figure=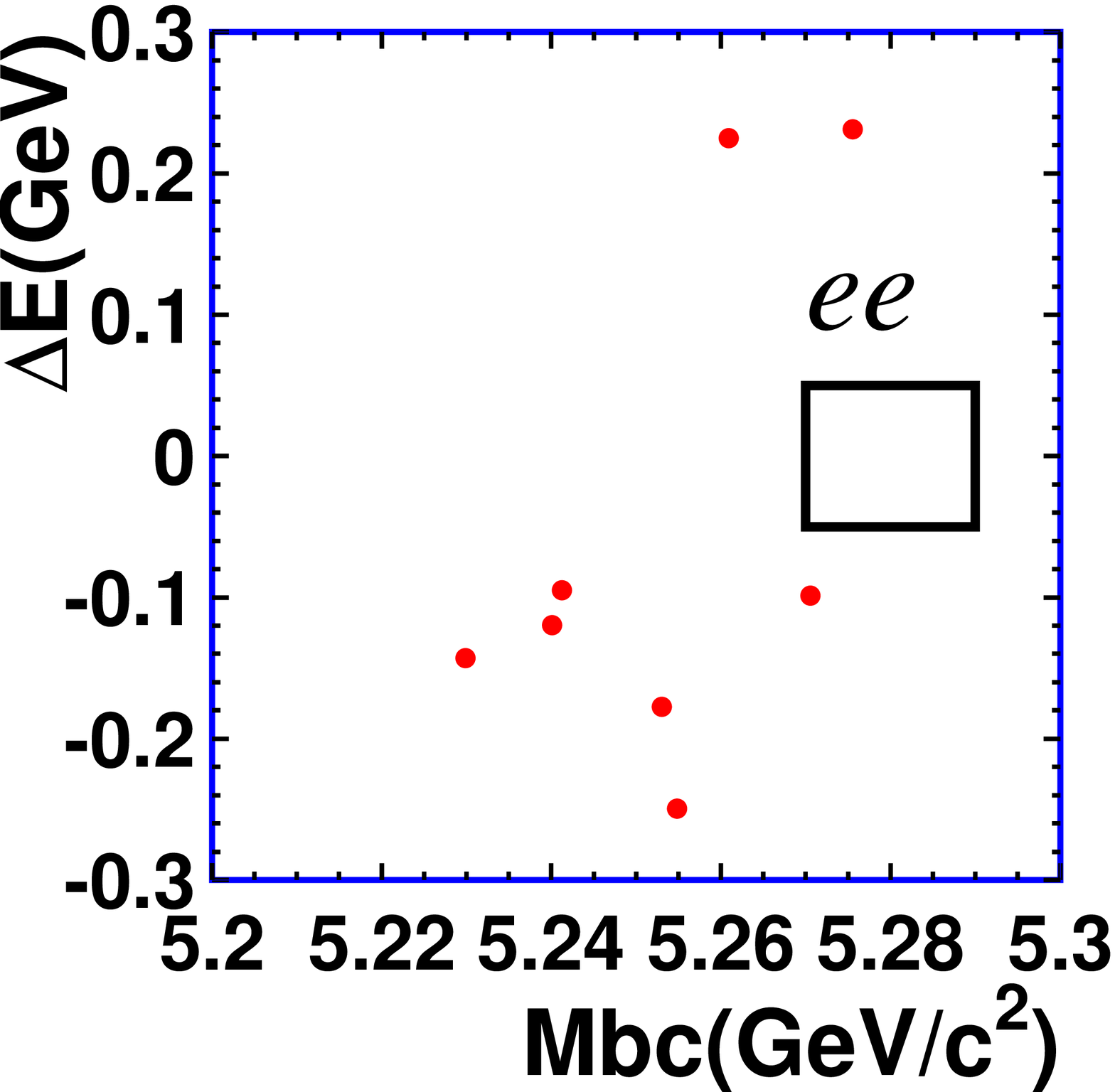,width=3truecm}
  \psfig{figure=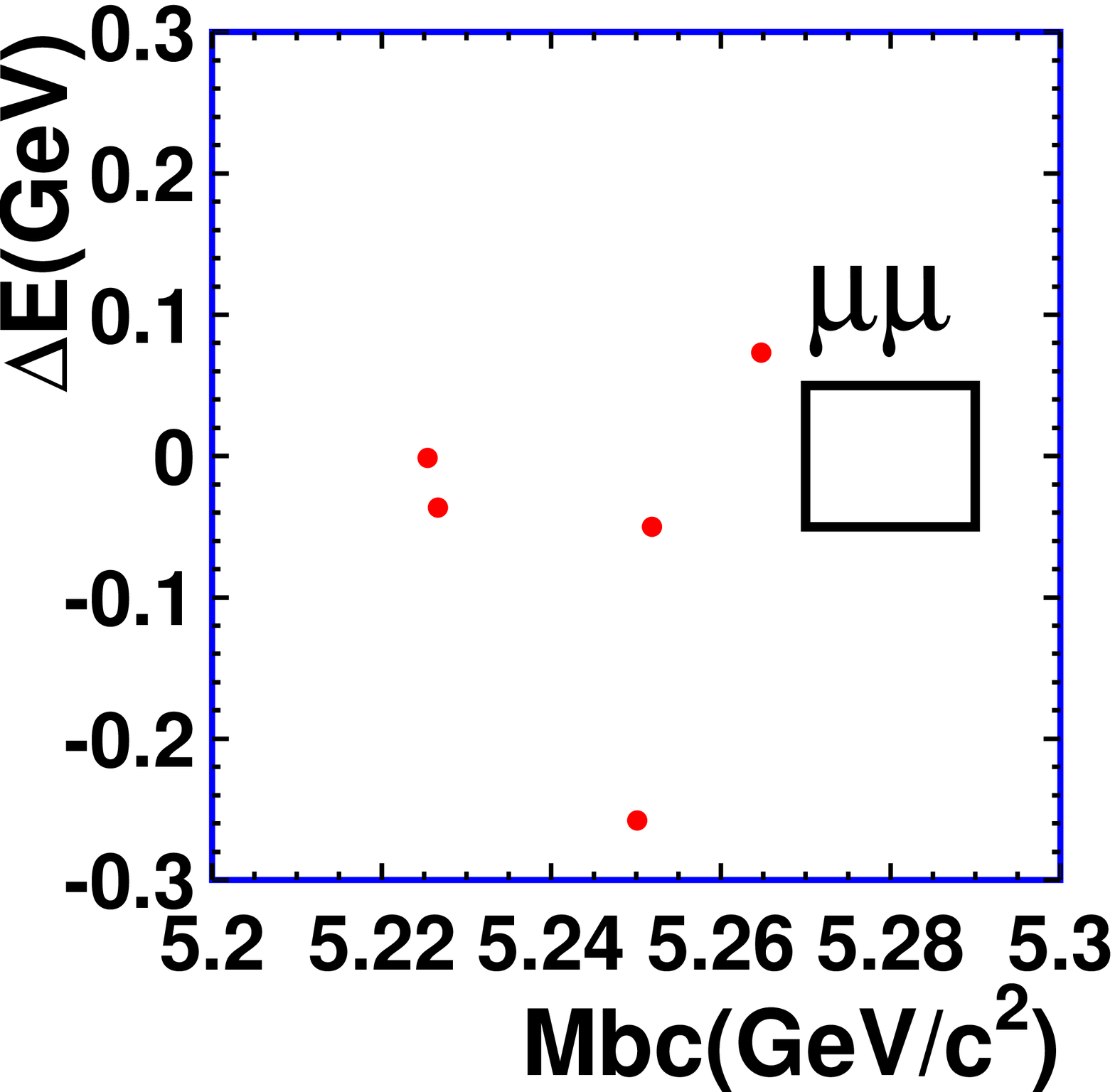,width=3truecm}
  \psfig{figure=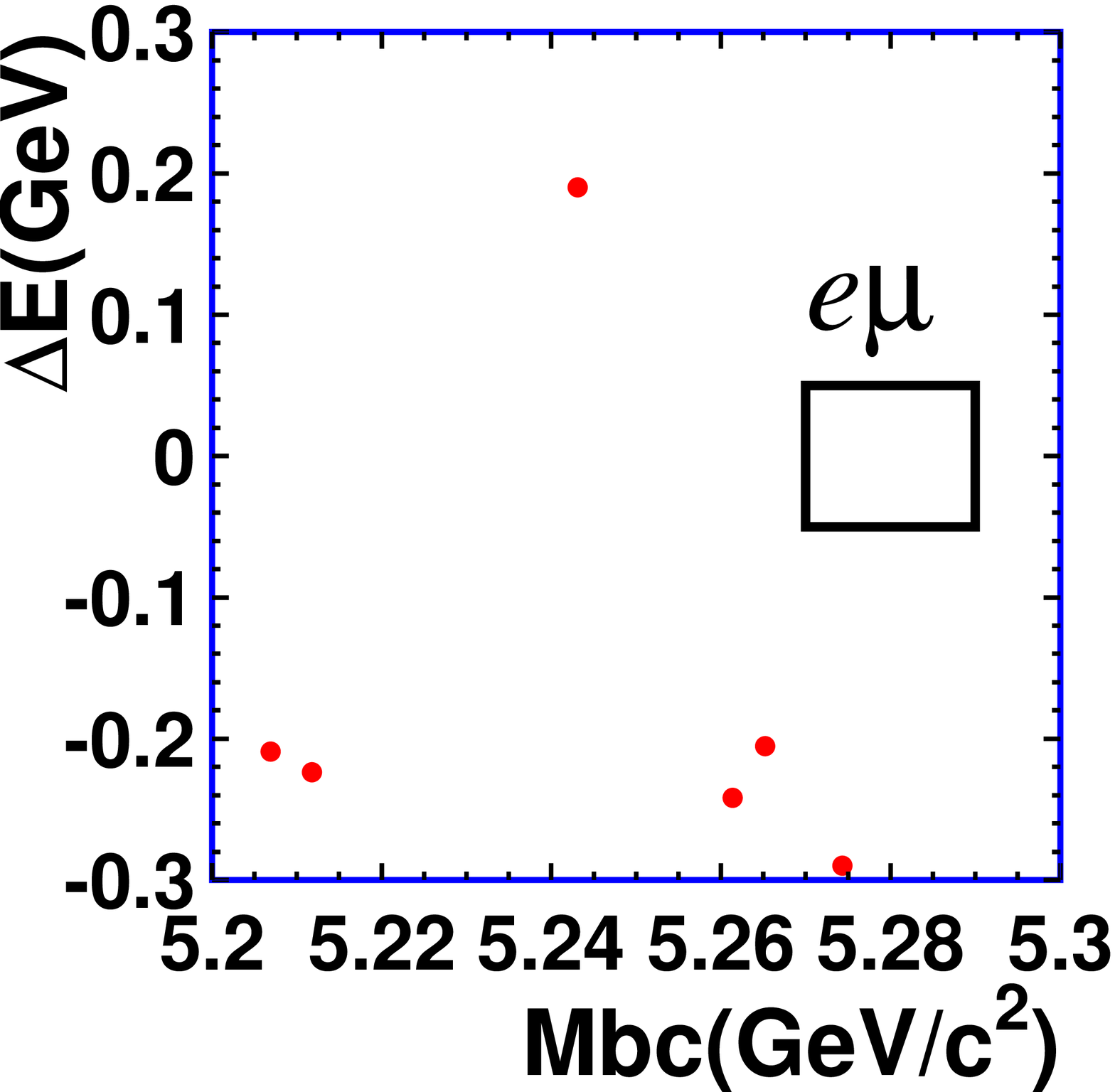,width=3truecm}
  \caption{$B_d^0\to \elel$ search results from Belle.}
  \label{fig:ll-belle}
\end{figure}

Belle has searched for the the decays $B_d^0\to\epem$, $B_d^0\to\mumu$
and $B_d^0\to e^\pm\mu^\mp$, using a $78\fbinv$ data
sample\CITE.{bib:ll-belle}  The analysis method is similar to those for
the other exclusive decays.  The dominant background source is the
continuum $\epem\to\ccbar$ production in which both charm quarks decay
into leptons.  Leptons from $\epem\to\tau^+\tau^-$ and two-photon
processes can be removed by requiring five or more charged tracks in a
event.  No event was observed (Fig.~\ref{fig:ll-belle}) for the expected
background events of 0.2 to 0.3, and the upper limits set are
\begin{equation}
\begin{array}{rcl}
\Br(B_d^0\to\epem)&<&1.9\EM7 \\
\Br(B_d^0\to\mumu)&<&1.6\EM7 \\
\Br(B_d^0\to e^\pm\mu^\mp)&<&1.7\EM7 \\
\end{array}
\end{equation}
at the 90\% confidence level.

For the $B_s^0$ decays, $113\pbinv$ and $100\pbinv$ of Run-II data from CDF and D0 respectively
have been analyzed.  Both analyses require three variables to
reduce backgrounds and search for the signal in the $\mumu$ mass
distribution.  CDF uses the proper lifetime $c\tau$, the direction
difference in azimuthal angle between the $\mumu$ vertex and momentum
directions $\Delta\Phi$, and a measure of isolation of the $B_s^0$
candidate based on the tracks inside the cone around the $B_s^0$
direction; D0 also uses a similar set of variables.

\begin{figure}
  \center
  \psfig{figure=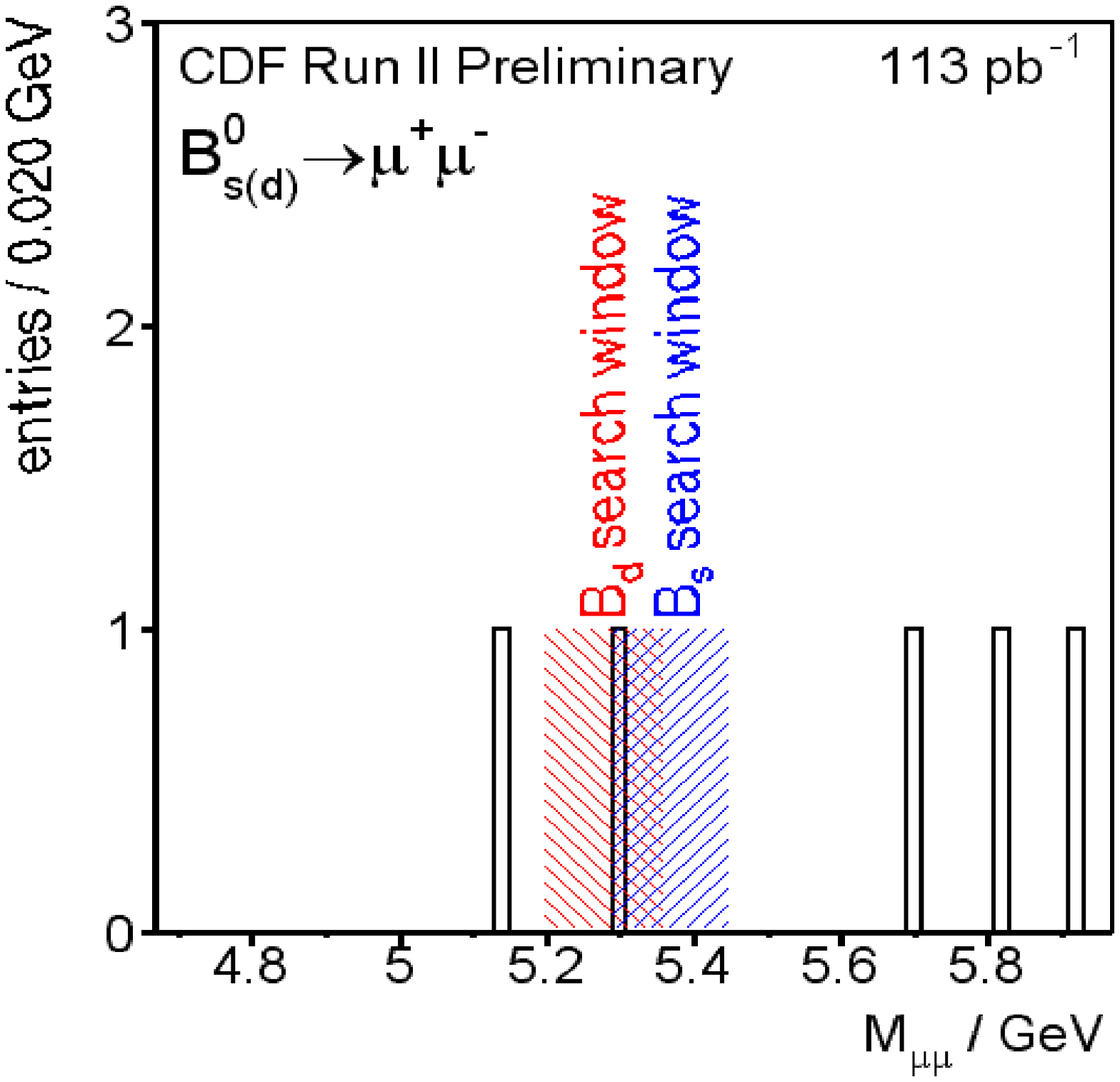,width=5.5truecm}
  \caption{$B_s^0\to \mumu$ search results from CDF.}
  \label{fig:bsmumu-cdf}
\end{figure}

\begin{figure}
  \center
  \psfig{figure=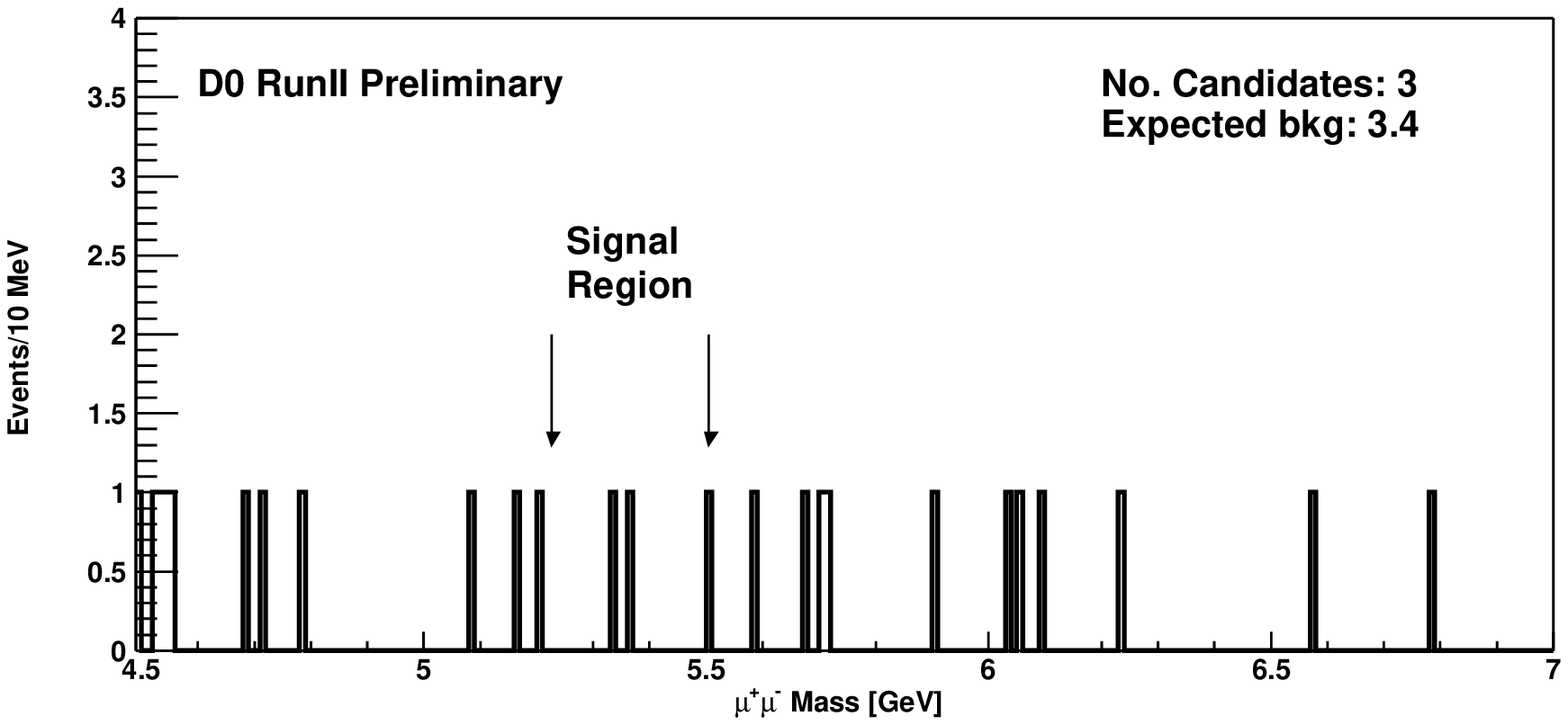,width=8truecm}
  \caption{$B_s^0\to \mumu$ search results from D0.}
  \label{fig:bsmumu-d0}
\end{figure}

CDF and D0 find one and three candidates as shown in
Figs.~\ref{fig:bsmumu-cdf} and \ref{fig:bsmumu-d0}, respectively.  The
CDF result leads to the upper limit of
\begin{equation}
\Br(B_s^0\to\mumu) < 9.5\EM7\mbox{~(90\%~CL)}
\end{equation}
that supersedes the previous CDF Run-I result.  D0's limit is
$\Br(B_s^0\to\mumu) < 16\EM7$.  CDF also reports
$\Br(B_d^0\to\mumu)<2.5\EM7$, which is already competitive with
Belle's result.

\begin{figure*}
  \center
  \psfig{figure=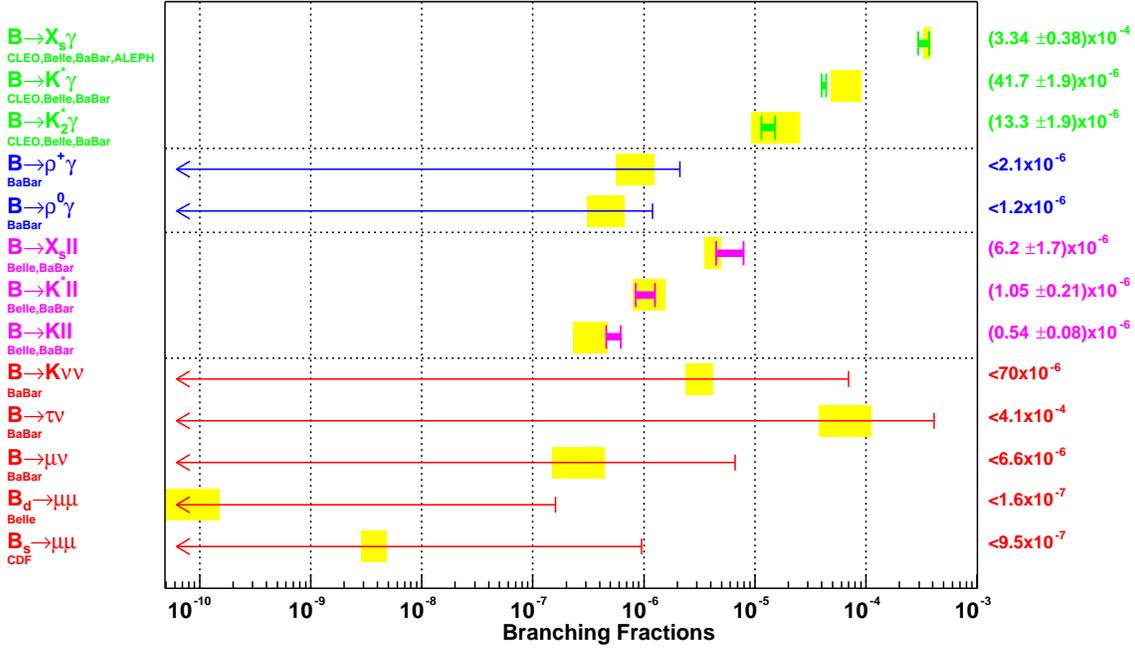,width=15truecm}
  \caption{Summary of branching fractions and upper limits compared with 
  the corresponding SM predictions.}
\vspace{-0.4cm}
  \label{fig:grand-summary}
\end{figure*}


\section{Conclusion}

Figure~\ref{fig:grand-summary} shows the currently measured branching
fractions and upper limits for the rare $B$ decays that involve a photon
or a lepton pair.  Most of the results have been updated rapidly along
with the accumulation of the $B$-factory data: new modes and
measurements in the exclusive $\btosgamma$ channels, new measurement of
the CP-asymmetry in $\BtoXsgamma$, the first observation of
$\BtoKstarll$ by Belle and evidence from BaBar, new results on $\BtoXsll$
from BaBar in agreement with Belle's, and new limits on $B\to K\nu\nubar$
and pure leptonic decays from BaBar, Belle and CDF are included.  So
far none of these results indicate a deviation from the SM.  As
$\BtoKstarll$ is finally measured, the next target will be the $b\to
d\gamma$ transition in the decay $\Btorhogamma$.

There are still many programs to be pursued using these already observed
rare $B$ decay channels, in addition to the searches for unobserved
modes.  One example is a measurement for mixing-induced CP-violation in
$\btosgamma$, for example with $B\to K^{*0}\gamma\to\KS\piZ\gamma$.
This channel has been considered to be experimentally challenging due to
the displaced $\KS$ decay vertex; however, BaBar has recently
demonstrated that it is possible to measure the $B$ decay vertex from
$\KS$ in the $B^0\to\KS\piZ$ channel\CITE,{bib:lp03-browder} and the
same technique is applicable to $\BtoKstargamma$.  The other example is
the measurement of the forward-backward asymmetry in $\BtoKstarll$ or
$\BtoXsll$.  These examples demand an order of magnitude larger data
sample than is available.  Fortunately, Belle and BaBar are still
collecting more data with improved luminosities expected, and are planning
to extend their luminosities by orders of magnitude.

\balance

\section*{Acknowledgments}
I wish to thank all the members of the Belle collaboration, Jeff
Richman, Stephane Willocq and Mark Convery for providing the latest
BaBar results, Rich Galik for the CLEO results, Majorie Shapiro for the
CDF results, Brad Abbott and Vivek Jain for the D0 results, and Paoti
Chang, Jim Alexander and Jim Smith of the Heavy Flavor Averaging Group
for the averaged numbers from the last minute results.  I acknowledge
Enrico Lunghi and Mikolaj Misiak among many theorists for many useful
private discussions.  I have to note that all the progress would not have 
been possible at all without excellent accelerator performances by the
KEKB and PEP-II accelerator teams.  Last but not least, I would like to
thank the Lepton-Photon '03 organizers for all their efforts in the
excellent conference organization.




\begin{thebibliography}{99}

\bibitem{bib:belle-phiks}
  Belle Collaboration, K. Abe \etal, arXiv:hep-ex/0308035.

\bibitem{bib:xsgam-cleo}
  CLEO  Collaboration, S. Chen \etal,   \Journal{\PRL}{87}{251807}{2001}.

\bibitem{bib:xsgam-aleph}
  ALEPH Collaboration, R. Barate \etal, \Journal{\PLB}{429}{169}{1998}.

\bibitem{bib:xsgam-belle}
  Belle Collaboration, K. Abe \etal,    \Journal{\PLB}{511}{151}{2001}.

\bibitem{bib:xsgam-babar-full}
  BaBar Collaboration, B.~Aubert \etal, arXiv:hep-ex/0207076.

\bibitem{bib:xsgam-babar-semi}
  BaBar Collaboration, B.~Aubert \etal, arXiv:hep-ex/0207074.

\bibitem{bib:xsgam-avg}
  C.~Jessop, SLAC-PUB-9610 (2002).

\bibitem{bib:xsgam-sm-latest}
  P.~Gambino and M.~Misiak, \Journal{\NPB}{611}{338}{2001}.

\bibitem{bib:xsgam-sm-nlo}
  K.~Chetyrkin, M.~Misiak and M.~M\"unz, \Journal{\PLB}{400}{206}{1997},
                             \Journal{Erratum ibid.\ B}{425}{414}{1998};
  A.~Kagan and M.~Neubert, \Journal{\EPJC}{7}{5}{1999}.


\bibitem{bib:xsgam-2hdm}
  F.~Borzumati, C.~Greub,   \Journal{\PRD}{58}{074004}{1998} 

\bibitem{bib:xsgam-bsm}
  For example,
  M.~Ciuchini, G.~Degrassi, P.~Gambino and G.~F.~Giudice, 
                            \Journal{\NPB}{534}{3}{1998};    
  C.~Bobeth, M.~Misiak and J.~Urban, 
                            \Journal{\NPB}{567}{153}{2000};  
  M.~Carena, D.~Garcia, U.~Nierste and C.~Wagner,
                            \Journal{\PLB}{499}{141}{2001}.  


\bibitem{bib:kstargam-cleo}
  CLEO Collaboration, T.~E.~Coan \etal, \Journal{\PRL}{84}{5283}{2000}.

\bibitem{bib:kstargam-babar}
  BaBar Collaboration, B.~Aubert \etal, \Journal{\PRL}{88}{101905}{2002}.

\bibitem{bib:kstargam-belle}
  Belle Collaboration, K.~Abe \etal, Belle-CONF-0319.

\bibitem{bib:kstargam-sm}
  S.~Bosch and G.~Buchalla, \Journal{\NPB}{621}{459}{2002};
  A.~Ali and A.~Parkhomenko, \Journal{\EPJC}{23}{89}{2002}.

\bibitem{bib:kstargam-becirevic}
  D.~Becirevic, arXiv:hep-ph/0211340; D.~Becirevic, talk given at 
  Ringberg Phenomenology Workshop on Heavy Flavors,
  Rottach-Egern, Germany, April 27 - May 2, 2003.


\bibitem{bib:kstargam-isospin}
  A.~Kagan and M.~Neubert, \Journal{\PLB}{539}{227}{2002}.


\bibitem{bib:m0gam-atwood}
  D.~Atwood, M.~Gronau and A.~Soni, \Journal{\PRL}{79}{185}{1997}.  

\bibitem{bib:kpipigam-gronau}
  M.~Gronau, Y.~Grossman, D.~Pirjol and A.~Ryd,
                                  \Journal{PRL}{88}{051802}{2002}.

\bibitem{bib:kxgam-belle}
  Belle Collaboration, S.~Nishida \etal, \Journal{PRL}{89}{231801}{2002}.

\bibitem{bib:k2gam-babar}
  BaBar Collaboration, B.~Aubert \etal, arXiv:hep-ex/0308021.

\bibitem{bib:veseli-olsson}
  S.~Veseli and M.~G.~Olsson, \Journal{\PLB}{367}{309}{1996}

\bibitem{bib:kphigam-belle}
  Belle Collaboration, A.~Drutskoy \etal, arXiv:hep-ex/0309006.

\bibitem{bib:Lpgam-cleo}
  CLEO Collaboration, K.~Edwards \etal, \Journal{\PRD}{68}{011102}{2003}


\bibitem{bib:acpxsgam-sm-orig}
  J.~Soares, \Journal{\NPB}{367}{575}{1991}.

\bibitem{bib:acpxsgam-kagan}
  A.~Kagan and M.~Neubert, \Journal{\PRD}{58}{094012}{1998}.

\bibitem{bib:acpxsgam-bsm}
  K.~Kiers, A.~Soni and G.~Wu, \Journal{\PRD}{62}{116004}{2000};
  S.~Baek and P.~Ko, \Journal{\PRL}{83}{488}{1998}.

\bibitem{bib:acpxsgam-cleo}
  CLEO Collaboration, T.~Coan \etal, \Journal{PRL}{86}{5661}{2001}.

\bibitem{bib:acpxsgam-belle}
  Belle Collaboration, K.~Abe \etal, arXiv:hep-ex/0308038.


\bibitem{bib:rhogam-babar}
  BaBar Collaboration, B.~Aubert \etal, arXiv:hep-ex/0306038.

\bibitem{bib:rhogam-belle}
  M. Nakao for the Belle Collaboration, talk given at 2nd Workshop on
  the CKM Unitarity Triangle, Durham, England, Apr.\ 2003,
  arXiv:hep-ex/0307031.

\bibitem{bib:rhogam-lunghi}
  T.~Hurth and E.~Lunghi, arXiv:hep-ex/0307142.


\bibitem{bib:xsll-sm}
  C.~Bobeth, M.~Misiak and J.~Urban, \Journal{\NPB}{574}{291}{2000};
  H.~H.~Asatrian, H.~M.~Asatrian, C.~Greub and M.~Walker,
                                     \Journal{\PLB}{507}{162}{2001};
  H.~H.~Asatrian, H.~M.~Asatrian, C.~Greub and M.~Walker,
                                     \Journal{\PRD}{65}{074004}{2002}.

 \bibitem{bib:xsll-sm-ali}                   
  A.~Ali, E.~Lunghi, C.~Greub and G.~Hiller, \Journal{PRD}{66}{034002}{2002}.

\bibitem{bib:xsll-bsm}
  For example,
  E.~Lunghi, A.~Masiero,~I.~Scimemi and L.~Silverstrini,
                                               \Journal{\NPB}{568}{120}{2000};
  J.~L.~Hewett and J.~D.~Wells,                \Journal{\PRD}{55}{5549}{1997};
  T.~Goto, Y.~Okada, Y.~Shimizu and M.~Tanaka, \Journal{\PRD}{55}{4273}{1997};
  G.~Burdman, \Journal{\PRD}{52}{6400}{1995};
  N.~G.~Deshpande, K.~Panose and J.~Trampeti\'c, 
                                               \Journal{\PLB}{308}{322}{1993};
  W.~S.~Hou, R.~S.~Willey and A.~Soni,         \Journal{\PRL}{58}{1608}{1987}.


\bibitem{bib:kll-belle}
  Belle Collaboration, K.~Abe \etal, \Journal{\PRL}{88}{021801}{2002}.

\bibitem{bib:kll-babar}
  BaBar Collaboration, B. Aubert {\it et al.}, arXiv:hep-ex/0207082.

\bibitem{bib:kstarll-belle}
  Belle Collaboration, A.~Ishikawa \etal, arXiv:hep-ex/0308044.

\bibitem{bib:kstarll-babar}
  BaBar Collaboration, B.~Aubert \etal, arXiv:hep-ex/0308042.

\bibitem{bib:kstarll-ali}
  A.~Ali, E.~Lunghi, C.~Greub and G.~Hiller, \Journal{\PRD}{66}{034002}{2002};
  E.~Lunghi, arXiv:hep-ph/0210379.

\bibitem{bib:kll-sm}
  For example,
  D.~Melikhov, N.~Nikitin and S.~Simula, \Journal{\PLB}{410}{290}{1997};
  P.~Colangelo, F. De Fazio, P. Santorelli and E.~Scrimieri, 
                             \Journal{\PRD}{53}{3672}{1996},
                             \Journal{Erratum-ibid. D}{57}{3186}{1998};
  M.~Zhong, Y.~L.~Wu and W.~Y.~Wang, 
                          \Journal{Int. J. Mod. Phys. A}{18}{1959}{2003};
  A.~Faessler \etal,                     \Journal{\EPJD}{4}{18}{2002};
  T.~M.~Aliev, C.~S.~Kim and Y.~G.~Kim,  \Journal{\PRD}{62}{014026}{2000};
  W.~Jaus and D.~Wyler,                  \Journal{\PRD}{41}{3405}{1990}.


\bibitem{bib:xsll-belle}
  Belle Collaboration, J.~Kaneko \etal, \Journal{\PRL}{90}{021801}{2003}.

\bibitem{bib:xsll-babar}
  BaBar Collaboration, B.~Aubert \etal, arXiv:hep-ex/0308016.


\bibitem{bib:knunu-semilep}
  BaBar Collaboration, B.~Aubert \etal, arXiv:hep-ex/0207069.

\bibitem{bib:knunu-hadronic}
  BaBar Collaboration, B.~Aubert \etal, arXiv:hep-ex/0304020.

\bibitem{bib:knunu-sm}
  G.~Buchalla, G.~Hiller and G.~Isidori, \Journal{\PRD}{63}{014015}{2001}.


\bibitem{bib:taunu-semilep}
  BaBar Collaboration, B.~Aubert \etal, arXiv:hep-ex/0303034.

\bibitem{bib:taunu-hadronic}
  BaBar Collaboration, B.~Aubert \etal, arXiv:hep-ex/0304030.

\bibitem{bib:taunu-l3}
  L3 Collaboration, M.~Acciarri \etal, \Journal{\PLB}{396}{327}{1997}.

\bibitem{bib:munu-belle}
  Belle Collaboration, K.~Abe \etal, Belle-CONF-0247.

\bibitem{bib:munu-babar}
  BaBar Collaboration, B.~Aubert \etal, arXiv:hep-ex/0307047.


\bibitem{bib:ll-sm}
  A.~Buras, \Journal{\PLB}{566}{115}{2003}.

\bibitem{bib:ll-belle}
  Belle Collaboration, M.-C.~Chang \etal, arXiv:hep-ex/0309069.


\bibitem{bib:lp03-browder}
  T.~Browder, talk given at XXI International Symposium on Lepton and
  Photon Interactions at High Energies, Batavia, Illinois, USA,
  Aug.\ 2003, in this Proceedings.

\end{thebibliography}
\end{document}